\documentclass[journal]{IEEEtran}

\usepackage{amsmath}
\usepackage{amssymb}
\usepackage{amsthm}
\usepackage{booktabs}
\usepackage{array}
\usepackage{graphicx}
\usepackage{tikz}
\usetikzlibrary{arrows.meta,positioning,fit}
\usepackage[hidelinks]{hyperref}
\usepackage{etoolbox}
\newtheorem{theorem}{Theorem}
\newtheorem{lemma}{Lemma}
\newtheorem{proposition}{Proposition}
\newtheorem{remark}{Remark}
\newtheorem{assumption}{Assumption}

\title{Call-Level Revision Contracts for LLM-Authored Workflows}

\author{Chenyang Jiang\textsuperscript{1}, Benyang Pan, and Puqu Wu\\[0.4ex]
\small\textsuperscript{1}South China Normal University, Guangzhou, China}

\hypersetup{
  pdftitle={Call-Level Revision Contracts for LLM-Authored Workflows},
  pdfauthor={Chenyang Jiang, Benyang Pan, Puqu Wu},
  pdfsubject={LLM-authored workflow repair, durable execution, idempotency},
  pdfkeywords={LLM agents, workflow repair, revision contract, durable execution, idempotency}
}

\begin{document}
\maketitle

\thispagestyle{plain}
\pagestyle{plain}

\begin{abstract}
\normalfont
\noindent Durable execution engines replay a recorded history and do not compare
activity payloads. When a model repairs a workflow that failed partway through, it
must either pay again for calls that succeeded or accept a recorded result whose
payload changed. We define a revision contract for workflows whose orchestration code
is written by a model. Each capability call carries a structural identity built from
its callsite, its fanout position, and its occurrence count, and the identity stays
stable under rewrites that leave the call in place, so committed identities form a frozen prefix. The frozen unit is the call expression, and the surrounding code stays
writable. A static gate rejects a rewrite that would change a committed call before a container
starts. A runtime digest stops a changed payload at a committed identity, and an
occurrence ledger names a repeated paid payload. Under three assumptions, a candidate the gate
accepts replays every committed call it invokes, executes only new work, and reports
the identities it skips. No local identity is both relocation-invariant and free of
wrong reuse, so callsite identity turns a relocation into a pre-spend stop. On media
workflows, 22 gate-passed repairs replay 258 committed identities with no duplicate,
and a live six-way fanout makes seven paid calls instead of twelve. A 12-task
customer-support study applies the same identity, replay, and boundary checks on two
models.

\end{abstract}


\section{Introduction}
\label{sec:intro}

A repaired workflow may add work, but it must not pay twice for work that already ran,
and it must not return a result that the repair has replaced. We make both properties an
explicit contract for workflows whose orchestration code is written by a model.

Tasks that used to be a single model call are now small programs. A request for a thirty-second brand video with six shots becomes a plan for a storyboard and one image call per shot. Each image is reviewed, the failures are regenerated, and the result is composed. Our platform lets the model write the orchestration script itself, in a restricted
Python subset with capability primitives such as
\verb|ctx.run("image.generate", payload)| and a broker that authorizes every call. The
script may loop, fan out, retry, and branch on data that earlier calls produced, which a
fixed DAG cannot express.

That flexibility has two failure modes, and both cost money. In the first, three shots
are generated and paid for and the fourth call fails. The natural repair is to show the
model the error and let it rewrite, and the rewritten script calls the provider for the
first three shots again. The second mode is the reverse and is worse for the product.
The model decides that a committed payload was wrong, for example a prompt, a size, or a
seed, and edits it in place. A replay engine does not compare payloads, so it returns
the recorded result and the run succeeds with the replaced artifact.

Neither gap is closed by a durable execution engine, which replays a recorded history,
checkpoints after each workflow task, and requires the replayed code to issue the same
commands in the same order. The guarantee is stated for the program that produced the
history. A repair that moves a call into a helper still replays, because the command
sequence is unchanged. A repair that edits a committed payload is served from the
history, because the payload is not part of the key. Prior work formalizes resume for a
fixed program \cite{Resume26}, semantic step identity \cite{Pollard26}, runtime revision
scope \cite{ControlScope26}, and replay capsules under an unchanged source
\cite{State26}. What a rewritten script may still touch is left open.

After a failure, the committed calls cannot change and the rest of the script can. The two sets are not separated by a line number, and under \verb|parallel| calls arrive in an
unstable order. In a loop, one source location executes many times, and formatting moves
lines. Each capability call therefore carries a structural identity
\(\iota=(\sigma,\pi,\omega)\), made of the callsite \(\sigma\) (enclosing function,
block path, and call ordinal), the fanout path \(\pi\), and the occurrence counter
\(\omega\). None of the three depends on the payload or on arrival order, so a
reordering cannot disguise a replay as a new call, or a new call as a replay.

The contract has three checks over a frozen prefix, and a bounded repair loop that uses them. Committed identities are durable, and a callsite with a
committed occurrence is frozen. Before a container starts, a static gate rejects a
candidate that removes a frozen callsite, changes its capability, or changes its call
expression. At dispatch, a payload digest stops a call that reaches a committed identity
with a different payload, so the run neither pays again nor returns a stale artifact. An occurrence ledger names a call at a new identity whose payload was already paid for at the same site and path. The repair loop returns structured counterexamples to the model, and a call-count check rejects a candidate that drops calls. Under the three assumptions of Section~\ref{sec:formal}, a candidate that the gate accepts never executes a committed identity.

The contract freezes the call expression and leaves the statements around it writable. A
loop may grow, a bound may change, and prompt assembly may be rewritten. Moving a
committed call into a helper is the harder case. The move preserves the capability, the
payload, and the order, but the moved call has a new callsite, so under structural
identity it is new work and would be paid for. The gate stops the candidate before any
provider call. Section~\ref{sec:background} walks through a real repair of this shape. Proposition~\ref{prop:dilemma} shows that no local identity can be both relocation-invariant and free of wrong reuse, so the choice is forced.

We evaluate the contract on media workflows and on a customer-support corpus, with a
live image provider, two durable engines, and two model families. On eight real failures
the model repairs every script within two rewrites, and the gate stops the one repair
that would have paid for a committed call again. The spend advantage is strict when a
failure falls inside a fanout, where a full rerun or an engine reset repeats the
committed branches, and it disappears on a sequential tail repair, which Section~\ref{sec:eval} reports.

Section~\ref{sec:background} describes durable execution, the platform, and the repair
that motivates the contract. Section~\ref{sec:formal} gives the model and the safety
result, Section~\ref{sec:design} the mechanisms and their boundary, and
Section~\ref{sec:impl} the implementation. Section~\ref{sec:eval} reports the
evaluation, Section~\ref{sec:threats} what it covers, and Section~\ref{sec:related}
positions the work.

\section{Background and Motivating Example}
\label{sec:background}

\subsection{Durable execution and replay}

A durable execution engine records every external call in a history while the workflow
runs. On recovery it re-executes the workflow function from the beginning, serves
completed calls from the history, and requires the replayed code to issue the same
commands in the same order. Checkpoints are taken per workflow task, so the unit the
engine protects is the program that produced the history. Temporal, DBOS, and Restate
choose different checkpoint units and different recovery paths
\cite{Temporal,DBOS25,Restate25}. In all three, a recorded call is matched by its
position in the command sequence, not by the source expression that produced it. A
version marker can declare a code change compatible, and that declaration is the part an
engine cannot check.

\subsection{The platform}

A job arrives with a request and an envelope that fixes the granted capabilities, the
token budget, and the concurrency limit. The host gives the model a skill document and a
capability catalog, and the model returns a Python script whose entry point is
\texttt{orchestrate(ctx, args)}. The only route to an external effect is
\texttt{ctx.run(name, payload)}. The context also exposes \texttt{parallel},
\texttt{pipeline}, \texttt{dag}, and \texttt{race} for fanout and ordering. A checker
rejects imports, dunder access, unbounded loops, and unguarded fanout before the script
runs, and a broker outside the container authorizes each capability call against the
grant set. The script executes in a networkless container with a read-only root
filesystem and no provider credentials. The run store keeps one record per committed
call.

Four capabilities are production-bound in this build. \texttt{image.generate} is a paid
image call. \texttt{storyboard.plan}, \texttt{review.visual}, and \texttt{text.generate}
are read-only model calls. \texttt{video.generate} is catalog-only and appears in
authoring studies, where no provider is billed.

\subsection{A repair that must be stopped}

Figure~\ref{fig:a425} shows a script that the model wrote for a four-shot scout run, and
the repair it produced after the fourth image call failed at the provider. The repair
factors the image call into a helper named \texttt{gen\_one} and calls the helper from
the same place. This is an ordinary refactor, because the capability multiset is unchanged, the
call order is unchanged, and every checker rule passes.

Under a command-sequence identity the repair replays, because the commands are the same.
Under the structural identity of Section~\ref{sec:formal} the committed callsite
\texttt{orchestrate/\#1} no longer exists, so the moved call is new work and would be
dispatched and paid for. The gate reports \texttt{removed} and stops the candidate
before the container starts. The stop is a design decision, and
Section~\ref{sec:design} states why no local identity avoids it.

\begin{figure}[t]
\centering
{\scriptsize
\begin{verbatim}
# committed: image.generate inside gen()
async def orchestrate(ctx, args):
    async def gen(s):
        return await ctx.run("image.generate",
                             {"prompt": s})
    return await ctx.parallel(
        [gen(s) for s in args["scout_shots"][:4]])

# repair: the call is moved into gen_one()
async def gen_one(ctx, s):
    return await ctx.run("image.generate",
                         {"prompt": s})

async def orchestrate(ctx, args):
    return await ctx.parallel(
        [gen_one(ctx, s) for s in args["scout_shots"][:4]])

# gate: removed  committed site orchestrate/#1
# spend: 0 provider calls; the repair never starts
\end{verbatim}
}
\caption{The a425 repair. Moving a committed call into a helper preserves the capability
multiset and the call order, so a checker-only repair accepts it. The frozen callsite
disappears, and the gate stops the candidate before spending.}
\label{fig:a425}
\end{figure}

\subsection{What the contract has to decide}

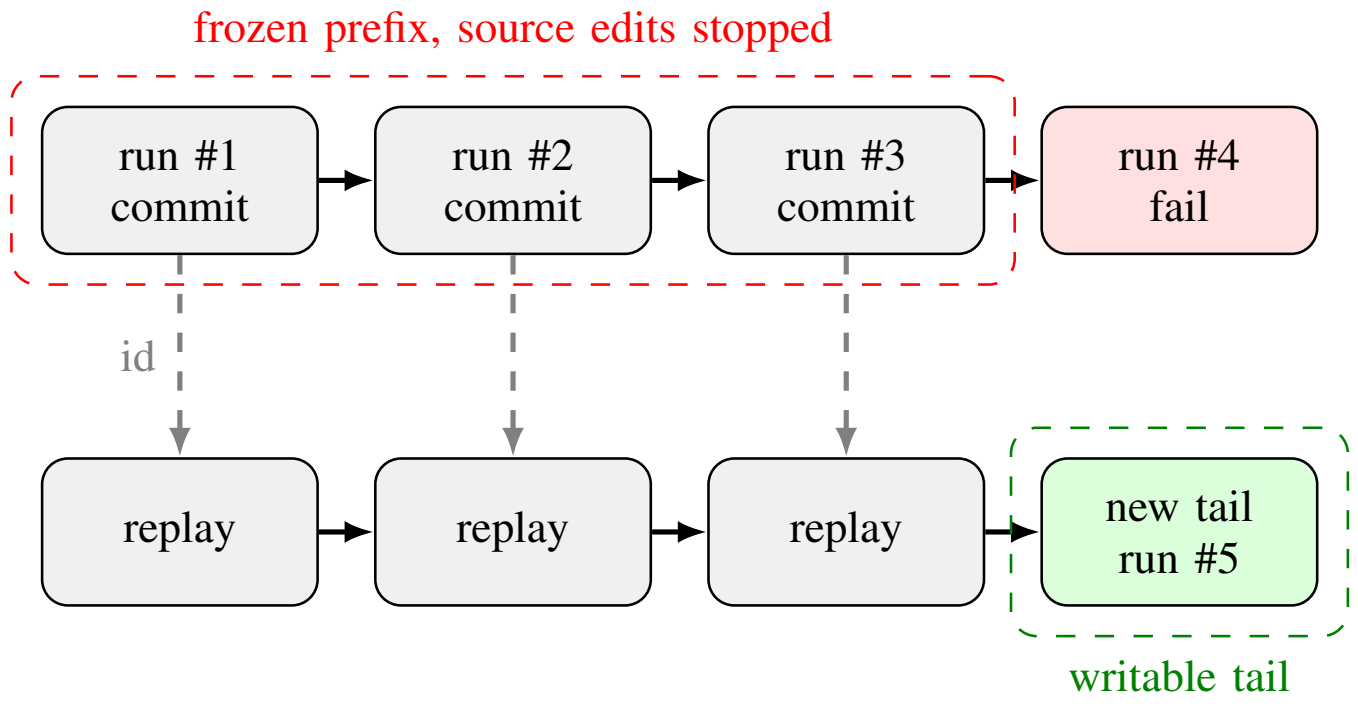
\begin{figure}[t]
\centering
\resizebox{\columnwidth}{!}{%
\begin{tikzpicture}[
  box/.style={draw, rounded corners, align=center, font=\scriptsize,
              minimum width=15mm, minimum height=8mm},
  arr/.style={-{Latex[length=1.6mm]}, thick},
  lab/.style={font=\scriptsize, align=center}
]
\node[box, fill=gray!12] (a1) {run \#1\\commit};
\node[box, fill=gray!12, right=3mm of a1] (a2) {run \#2\\commit};
\node[box, fill=gray!12, right=3mm of a2] (a3) {run \#3\\commit};
\node[box, fill=red!12, right=3mm of a3] (a4) {run \#4\\fail};
\node[box, fill=gray!12, below=11mm of a1] (b1) {replay};
\node[box, fill=gray!12, right=3mm of b1] (b2) {replay};
\node[box, fill=gray!12, right=3mm of b2] (b3) {replay};
\node[box, fill=green!14, right=3mm of b3] (b4) {new tail\\run \#5};
\draw[arr] (a1) -- (a2);
\draw[arr] (a2) -- (a3);
\draw[arr] (a3) -- (a4);
\draw[arr] (b1) -- (b2);
\draw[arr] (b2) -- (b3);
\draw[arr] (b3) -- (b4);
\draw[arr, dashed, gray] (a1) -- node[lab, left] {id} (b1);
\draw[arr, dashed, gray] (a2) -- (b2);
\draw[arr, dashed, gray] (a3) -- (b3);
\node[draw, dashed, red, rounded corners, fit=(a1)(a3), inner sep=1.6mm,
      label={[red, font=\scriptsize]above:frozen prefix, source edits stopped}] {};
\node[draw, dashed, green!50!black, rounded corners, fit=(b4), inner sep=1.6mm,
      label={[green!50!black, font=\scriptsize]below:writable tail}] {};
\end{tikzpicture}%
}
\caption{The revision boundary. The original run commits identities 1--3 and
fails at 4. A repair replays 1--3 (same structural identities), may append a new
call at a new identity, and may not edit the frozen prefix. The static gate stops such edits before a container starts. A control-flow edit that leaves
1--3 in the source but unreachable passes the gate and is reported as an orphan,
which is why fidelity is separate from safety.}
\label{fig:boundary}
\end{figure}

Figure~\ref{fig:boundary} shows the boundary the design has to draw. The original run
commits three calls and fails at the fourth. A repair may replay the three committed
calls and append work after them, but it may not edit the committed calls in place. The
figure also shows the case that separates safety from fidelity. A control-flow edit can
leave a committed callsite in the source and make it unreachable, which the static gate
accepts and the orphan report records.

Three requirements follow. A candidate that the gate accepts must not execute a
committed identity, because a second execution is a second payment. A committed identity
that the candidate reaches must replay or stop the run, because a changed payload cannot
be served from the record. The decision must be made before the container starts, since
a check that runs at dispatch can only detect the payment after it happened. The rest of
the paper asks which identity makes the first requirement checkable on the source, which
mechanisms enforce it at run time, and what no identity can achieve.

Three assumptions support the results. Every external effect goes through
\texttt{run}(c,p). Committed records are durable and only grow. The identity is
structural and does not depend on the payload, the arrival order, or the line number.
Section~\ref{sec:formal} states each assumption formally and gives the case that fails
when it does not hold.

\section{Model and Problem Statement}
\label{sec:formal}

Identity-level replay safety rests on three assumptions, and exactly one of them constrains
how the source may be revised. This section defines the execution model and the structural
identity of a capability call. It then states the safety theorem, which covers the committed
calls that a candidate keeps, and a separate fidelity condition that covers the calls it
skips.

\paragraph{Terminology}
A \emph{callsite} \(\sigma\) is a normalized AST position. An \emph{identity} \(\iota\) adds
the fanout path and the occurrence counter. A \emph{record} is what a committed identity
stores, and the store key is a \emph{slot}. The \emph{committed prefix} \(P\) is the set of
identities with records, and a callsite with a record is \emph{frozen}. A \emph{divergence}
is a call at a committed identity with a different payload, and it stops the run. An
\emph{orphan} is a frozen identity that the candidate does not invoke, and it is recorded
without failing the run.

\subsection{Scripts, callsites, and records}

A workflow script is an asynchronous function
\(S:\mathcal{X}\to\mathcal{Y}\), written in a restricted Python subset. Every external side
effect is mediated by \(\mathsf{run}(c,p)\), where \(c\) is a capability name and \(p\) is a
payload. The checker rejects imports, dunder access, and other ways around this surface, and
Section~\ref{sec:impl} maps the rules to the code.

A \emph{structural callsite} is
\begin{equation}
\sigma=(f,b_1,\ldots,b_k,j)\in\Sigma,
\end{equation}
where \(f\) is the enclosing function, each \(b_i\) is the kind and ordinal of a statement
block, and \(j\) is the ordinal of the \(\mathsf{run}\) call within the innermost block. The
identity is computed on a normalized AST, so comments, formatting, and line breaks do not
change \(\sigma\).

A container primitive \(\mathsf{parallel}\), \(\mathsf{pipeline}\), \(\mathsf{dag}\), or
\(\mathsf{race}\) pushes a branch index when it enters a branch. The resulting \emph{fanout
path} is
\begin{equation}
\pi=(i_1,\ldots,i_d)\in\Pi,\qquad \pi=()\ \text{for sequential code}.
\end{equation}
A site may execute several times in a loop, and \(\omega\in\mathbb{N}_0\) is the zero-based
occurrence counter for the pair \((\sigma,\pi)\). The \emph{call identity} is
\begin{equation}
\iota=(\sigma,\pi,\omega)\in\mathcal{I},
\end{equation}
and a \emph{record} is
\begin{equation}
r=(\iota,c,h,v,u),
\end{equation}
where \(h=H(\mathsf{canonical}(p))\) is the payload digest, \(v\) is the result reference,
and \(u\) records usage and cost. A run produces a trace \(T=\langle r_1,\ldots,r_n\rangle\).
The committed prefix \(P\) is the persisted partial map from \(\iota\) to \(r\).
\(\mathsf{commit}\) is idempotent on \(\iota\), and \(P\) only grows.

A frozen site constrains the \emph{call expression} rather than the block. A loop that
committed \(\omega=0,\ldots,k-1\) may execute \(\omega=k\), because that identity is new. The
expression \(\mathsf{run}(c,p)\) that produced the committed payloads may not change, while
the statements around it stay writable. A line-number heuristic and a block-level freeze
give different writable sets (Figure~\ref{fig:boundary}).

\subsection{Runtime rule}

For a call \((\iota,c,p)\), the runtime applies the following rule.
\begin{enumerate}
\item If \(P(\iota)=r\) and \(r.c=c\) and \(r.h=H(p)\), return \(r.v\). This is a replay
hit, not a provider execution.
\item If \(P(\iota)=r\) but \(r.c\ne c\) or \(r.h\ne H(p)\), raise
\(\mathsf{Divergence}(\iota)\). Continuing would place a new effect on top of an already
committed one at the same identity.
\item If \(\iota\notin P\), execute \((c,p)\), commit \((\iota,c,H(p),v)\), and return \(v\).
\end{enumerate}

The third branch is the one that spends money. The second raises a safety event, and the
first serves the past without re-executing it.

\subsection{The repair problem}

A repair replaces the failed script \(S\) with a candidate \(S'\) for the next attempt. The contract makes three requirements. Every identity in \(P\) that \(S'\) invokes is served from \(P\) or raises \(\mathsf{Divergence}\). Identities outside \(P\) may execute, so a repair can add work after the failure.
Every identity in \(P\) that \(S'\) skips is reported as an orphan. The gate
(Section~\ref{sec:gate}) decides the structural part before a container starts, the
runtime rule enforces the rest at dispatch, and the orphan report names what the
candidate skipped.

\subsection{Assumptions}

\begin{assumption}[Mediation]
\label{a:mediation}
Every external side effect goes through \(\mathsf{run}(c,p)\). The script reaches external
effects only through the runtime, so imports, file I/O, network access, clocks, and
randomness are outside the surface. The checker rejects the source-level routes
(Section~\ref{sec:impl}), and the sandbox removes the rest. A rule that the checker misses
therefore stays inside the runtime boundary, because the container has no network, a
read-only root filesystem, and no provider credentials.
\end{assumption}

\begin{assumption}[Durable monotone prefix]
\label{a:durable}
\(P\) is keyed by \(\iota\), \(\mathsf{commit}\) is idempotent on \(\iota\), and committed
records survive failures, container restarts, and source rewrites. Silent deletion or
overwrite is outside the trust model. The assumption requires a store that answers, and
Section~\ref{sec:system} states the fail-closed policy for an outage.
\end{assumption}

\begin{assumption}[Structural identity]
\label{a:identity}
The identity \(\iota=(\sigma,\pi,\omega)\) is determined by the normalized AST position, the
fanout path, and the occurrence counter. It does not depend on payload, arrival order,
wall-clock time, or line number. Formatting, comments, payload values, and appends
\emph{after} committed sites do not change a committed \(\sigma\). Different call instances
in one trace receive different \(\iota\). An insertion or deletion of a capability call
\emph{before} a committed site shifts \(\sigma\). The gate compares the frozen sites against
the candidate, and Section~\ref{sec:boundary} states how the resulting identities are
treated.
\end{assumption}

Assumption~\ref{a:identity} does not require two semantically identical calls to share an
identity. A new site with the same provider request is a new identity. The
semantic-duplication boundary of Section~\ref{sec:boundary} covers that case, and
Theorem~\ref{thm:safety} covers the identities that a gate-accepted candidate invokes.

\subsection{Identity stability}

The proof rests on one source-level property.

\begin{lemma}[Identity stability under tail revision]
\label{lem:stability}
Let \(S\) be a script and \(\sigma\) a callsite in \(S\). For any candidate
\(S'\) obtained from \(S\) by (i) changing comments, formatting, or blank lines,
(ii) changing the \emph{value} of a payload while preserving the call's AST shape,
(iii) inserting a statement that contains no capability call \emph{before} a
committed site, or (iv) appending statements \emph{after} the last committed
callsite, every committed runtime identity \((\sigma,\pi,\omega)\) has the same
value in \(S'\) as in \(S\).
\end{lemma}

\begin{proof}
By induction over the block traversal computed by \texttt{assign\_sites}. Comments and
formatting are not AST nodes. A payload value change leaves the call position untouched. A
non-capability statement does not increment the call ordinal, and an append adds statements
after the site.
\end{proof}

Two edits fall outside the lemma. Inserting a block around a committed site adds a label to
every nested call, so the old identifier is no longer produced and the frozen-site comparison
reports \texttt{removed}. Inserting a fanout branch or a loop iteration before a committed
identity changes \(\pi\) or \(\omega\). The static gate cannot see either shift, and
Section~\ref{sec:boundary} states which runtime check catches it.

Lemma~\ref{lem:stability} constrains \(\sigma\), and the call-expression fingerprint is a
separate check. A literal payload edit preserves \(\sigma\) but changes the fingerprint, so
the static gate stops it. A data-dependent payload edit preserves both and is caught by the
runtime digest at dispatch. The gate catches source-level drift, and the digest catches
value-level drift.

\subsection{Identity-level safety}

The runtime rule is a dispatch invariant that holds for every candidate, including one that
renames every site and thereby avoids \(P\). The theorem lifts it to the system. The gate
provides the structural premise before spending, and the orphan report provides the fidelity
signal.

\begin{theorem}[Identity-level once]
\label{thm:safety}
Under Assumptions~\ref{a:mediation}--\ref{a:identity}, let \(S'\) be a candidate that the
gate accepts. Then no identity in \(P\) reaches the execute branch. Every committed identity
that \(S'\) invokes is served from \(P\), or raises \(\mathsf{Divergence}\) if its payload
changed. Every committed identity that \(S'\) does not invoke is reported as an orphan.
\end{theorem}

\begin{proof}
Gate acceptance gives presence, capability, and call-expression fingerprint for every frozen
site (Section~\ref{sec:gate}). Assumption~\ref{a:durable} keeps the records readable, and
Assumption~\ref{a:identity} with Lemma~\ref{lem:stability} gives the same identity at
dispatch. The runtime rule sends an invoked \(\iota\in P\) to the replay branch when its
digest matches, and to \(\mathsf{Divergence}\) when it does not. The execute branch requires
\(\iota\notin P\). The host compares the frozen identities with those invoked and reports
the difference as \(\mathsf{Orphan}\).
\end{proof}

Theorem~\ref{thm:safety} states the system-level guarantee, which is identity-level once with
an explicit fidelity signal. The proof separates four obligations. Identity stability
(Lemma~\ref{lem:stability}) is the source-level one. The gate establishes the structural part
before spending. The digest checks payload changes at dispatch. The orphan report names the
committed identities that a candidate skips.

\begin{remark}[Orphans are the fidelity signal]
\label{rem:orphan}
Gate acceptance preserves the committed callsites, and reachability is the candidate's choice.
The orphan report names every committed identity that a candidate skips, and
\texttt{DYNAMIC\_ORPHAN\_POLICY=stop} turns that report into a failed run. The default
\texttt{warn} records the orphan and lets the run complete, because an orphan is a fidelity
event rather than a second dispatch. A deployment that bills per artifact chooses
\texttt{stop}, and the default keeps the run auditable.
\end{remark}

\subsection{Replay fidelity}

\begin{proposition}[Replay fidelity]
\label{prop:fidelity}
If (a) each committed site's capability and call expression are unchanged, (b)
prefix control flow is deterministic under the same inputs and environment, (c) no
insertion or deletion before a committed site shifts its structural identity, and
(d) the candidate's control flow on the committed prefix is trace-equivalent to the
committed run (same branch decisions, loop counts, and fanout widths), then replay serves every identity in \(P\) exactly once, executes only identities outside
\(P\), and produces no orphan.
\end{proposition}

\begin{proof}[Proof sketch]
Induct over control steps. By (d), the candidate follows the committed trace on the
prefix, so every frozen site is reached in the same order and with the same loop
and fanout structure. By (c), the sequence of \((\sigma,\pi,\omega)\) triples is the
same as at commit time. By (a), each capability and payload digest matches its
record, so every \(\iota\in P\) takes the replay-hit branch. By (b), a second replay
of the same candidate makes the same decisions. After the prefix is replayed, remaining identities are outside \(P\) and enter the execute branch.
\end{proof}

The five script-level conditions of the checker approximate (a) and (b) syntactically.
They are ordered iteration over sets, statically bounded loops, no hidden inputs,
deterministic fanout width, and capability names within the grant envelope. Condition (d) is
a runtime property, and the orphan report names the identities that a candidate skips
(Remark~\ref{rem:orphan}).

\section{The Revision Contract}
\label{sec:design}

\subsection{Overview}

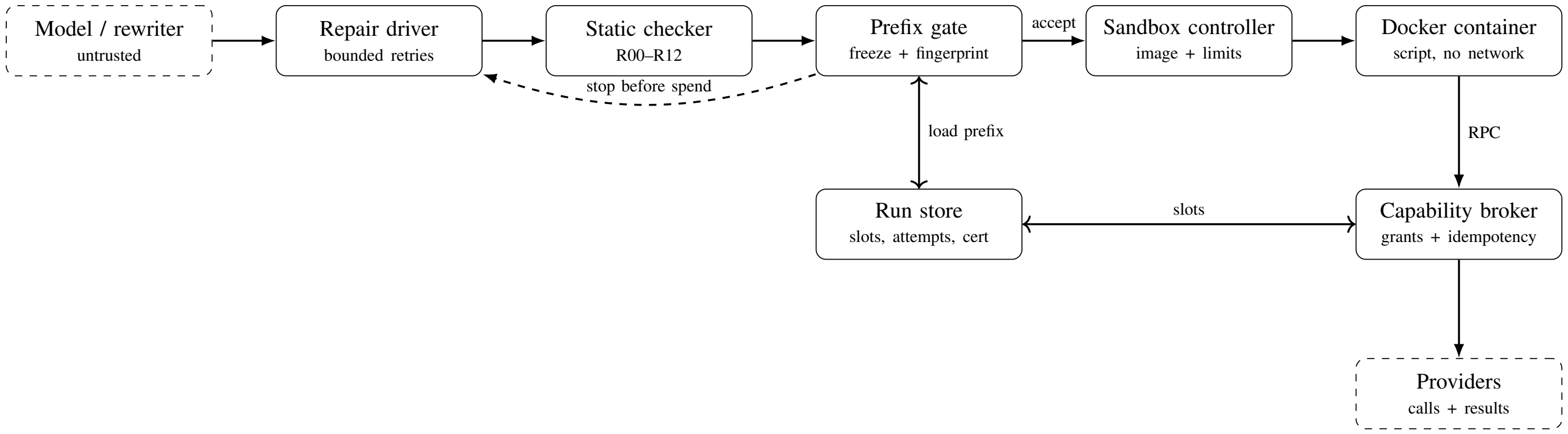
\begin{figure*}[t]
\centering
\resizebox{\textwidth}{!}{%
\begin{tikzpicture}[
  box/.style={draw, rounded corners, align=center, font=\small,
              minimum height=10mm, text width=2.7cm},
  ext/.style={draw, dashed, rounded corners, align=center, font=\small,
              minimum height=10mm, text width=2.7cm},
  arr/.style={-{Latex[length=2mm]}, thick},
  lbl/.style={font=\scriptsize, align=center}
]
\node[ext] (model) {Model / rewriter\\\scriptsize untrusted};
\node[box, right=9mm of model] (repair) {Repair driver\\\scriptsize bounded retries};
\node[box, right=9mm of repair] (check) {Static checker\\\scriptsize R00--R12};
\node[box, right=9mm of check] (gate) {Prefix gate\\\scriptsize freeze + fingerprint};
\node[box, right=9mm of gate] (sandbox) {Sandbox controller\\\scriptsize image + limits};
\node[box, right=9mm of sandbox] (container) {Docker container\\\scriptsize script, no network};
\node[box, below=16mm of gate] (store) {Run store\\\scriptsize slots, attempts, cert};
\node[box, below=16mm of container] (broker) {Capability broker\\\scriptsize grants + idempotency};
\node[ext, below=14mm of broker] (provider) {Providers\\\scriptsize calls + results};

\draw[arr] (model) -- (repair);
\draw[arr] (repair) -- (check);
\draw[arr] (check) -- (gate);
\draw[arr] (gate) -- node[lbl, above] {accept} (sandbox);
\draw[arr] (sandbox) -- (container);
\draw[arr] (container) -- node[lbl, right] {RPC} (broker);
\draw[arr] (broker) -- (provider);
\draw[arr, <->] (gate) -- node[lbl, right] {load prefix} (store);
\draw[arr, <->] (broker) -- node[lbl, above] {slots} (store);
\draw[arr, dashed] (gate) to[bend left=18] node[lbl, above] {stop before spend} (repair);
\end{tikzpicture}%
}
\caption{Host-side architecture. The model and the providers are external. The
script executes in a networkless, read-only container. The host annotates
callsites and owns the prefix, the broker, and the store.}
\label{fig:arch}
\end{figure*}

The host-side path appears in Figure~\ref{fig:arch}. The model writes an orchestration script, and the repair driver validates every candidate against the checker and the committed prefix. An accepted candidate is annotated with callsite identities and handed to the
sandbox, and capability calls return to the host through a bounded file RPC, pass through the
broker, and are committed to the run store. The container has no network and never holds a
provider credential. Three mechanisms catch the three ways in which a rewrite can repeat a
paid effect. A static gate protects the source-level call expression, and a runtime digest protects the payload at a committed identity. An occurrence ledger names a repeat that the first two cannot see.

\subsection{The frozen prefix and the static gate}
\label{sec:gate}

Let \(\mathcal{F}\) be the set of frozen sites. The gate accepts a candidate \(S'\) only if
every \(\sigma\in\mathcal{F}\) is present, keeps its capability name, and keeps its
call-expression fingerprint. A stopped candidate never reaches the container, so it spends
nothing. Acceptance covers presence, capability, and the call expression. Reachability and
the payload digest are separate checks, because neither can be decided from the source.

The host derives the frozen prefix from the committed slot records and builds a hash chain
over entries ordered by $(site, path, occurrence)$. The certificate carries the frozen-site
map, the chain digest, and the entry count. Verification recomputes both from the current
store and reports any difference, so the certificate is a consistency manifest against the
store.

The gate is an over-approximation of the preservation premise, and it is stricter than the premise requires. Proposition~\ref{prop:gate-sound} states the guarantee,
and Proposition~\ref{prop:gate-incomplete} states the cost.

\begin{proposition}[Gate soundness]
\label{prop:gate-sound}
If the gate accepts \(S'\), every committed callsite is present in \(S'\) with the same
capability and the same call-expression fingerprint, and the structural premise of
Theorem~\ref{thm:safety} holds. For every identity the candidate invokes, the runtime rule
gives identity-level once, and every identity it skips appears in the orphan report.
\end{proposition}

\begin{proof}
The gate checks presence, capability, and call-expression fingerprint
(Section~\ref{sec:gate}), which are the structural conjuncts of the preservation premise.
Reachability is checked at run time by the orphan report. When the identity is invoked, the
runtime rule applies Theorem~\ref{thm:safety}.
\end{proof}

\begin{proposition}[Pre-spend strictness]
\label{prop:gate-incomplete}
There is a candidate \(S'\) that preserves every committed call's capability, payload, and
order, and the gate reports it as \texttt{removed} and stops it before it starts. Relocating
a committed call from its enclosing function into a helper is such an edit. Under structural
identity the stop is necessary, because the moved call has a new \(\sigma\) and the runtime
would execute it and pay again.
\end{proposition}

\begin{proof}
Let \(\sigma=(f,b_1,\ldots,b_k,j)\) be a committed callsite. Move the call into a new helper
\(g\) invoked from the same point. The new site is \(\sigma'=(g,b'_1,\ldots,b'_{k'},j')\).
The capability and the call expression are unchanged, and the call is performed once, in the
same order. The runtime rule keys on \(\iota=(\sigma,\pi,\omega)\), and \(\sigma'\) is not in
\(P\), so the call would execute. The gate stops the candidate before spending and prevents
the duplicate. A command-sequence engine accepts the edit because its identity is the command
position, and the two identities disagree.
\end{proof}

The relocation case is the one repair among the four replay points of
Section~\ref{sec:eval} that the gate stops. A command-sequence engine accepts the edit and
replays the recorded result, and Section~\ref{sec:baselines} measures the difference.
Proposition~\ref{prop:dilemma} shows that the stop does not depend on the implementation.

\subsection{The runtime payload digest}

The gate reads the source, so it sees a literal payload edit and stops it. A payload that
depends on earlier results, such as a size read from the storyboard, preserves both the
callsite and the shape of the call expression. At dispatch the runtime hashes the canonical
payload and compares it with the record. A match replays \(r.v\), and a mismatch raises
\textsf{Divergence} before any new effect is recorded, so the run stops instead of paying
again or returning the replaced artifact. The digest is the only check that sees a value
computed at run time.

\subsection{The occurrence ledger}

The gate is static, and \(\pi\) and \(\omega\) are run-time quantities
(Section~\ref{sec:boundary}). The broker therefore keeps an occurrence ledger built from the
same slot records. For each \((\sigma,\pi)\) it indexes the committed payload digests. Before
a call that misses on identity is dispatched, the ledger asks whether the same site and path
already paid for this exact payload, and a hit is a \emph{semantic repeat}. Policies are
\(\mathsf{off}\), \(\mathsf{warn}\), \(\mathsf{deny}\), and the default \(\mathsf{auto}\),
which resolves per capability. The default denies a repeat for a side-effecting capability
and warns for a read-only one, so a repeated paid image call needs an explicit decision while
a repeated review is only budget. An operator who needs a permissive policy can set
\(\mathsf{warn}\) globally, and a repair that legitimately produces a second take declares
it.

The ledger does not stop a call on its own. It names the repeat and leaves the choice to
policy, which makes the occurrence-extension case visible, and optionally fatal, before money
is spent.

\subsection{The boundary of local identity}
\label{sec:identity-boundary}

The guarantee is keyed to the identity, and a pre-spend check can see the local syntactic context
of a call, which is the capability, the payload, the call expression, the fanout path, and
the occurrence, without the model's cooperation. Proposition~\ref{prop:dilemma} states the limit of that information.

\begin{proposition}[Identity boundary]
\label{prop:dilemma}
No local identity is both \emph{relocation-invariant} and free of \emph{wrong reuse} on the
candidates that keep a committed call's capability, payload, and order. Relocation
invariance means that moving the call into a new function or block preserves the identity,
and freedom from wrong reuse means that no call receives a result committed at a different
callsite.
\end{proposition}

\begin{proof}
Suppose \(\iota\) is local and relocation-invariant. A move into a helper preserves the
capability, the payload, the call expression, the fanout path, and the occurrence, so
\(\iota\) is unchanged. Two callsites can share a local context, and the same call in two
branches is the common case. They then share \(\iota\), so a record committed at the first is
a replay hit at the second, which is wrong reuse. If two callsites with the same local
context receive different identities, the difference is positional and depends on the
enclosing function or the block path. A move changes that position, so \(\iota\) changes, and
the gate stops the relocation or the runtime executes the call again. A local identity has no
third option.
\end{proof}

Table~\ref{tab:identity-choice} lists the alternatives. Three identities lie outside the
local class, and each makes a different trade. A \emph{global call ordinal} survives
relocation in sequential code, but an insertion before a committed call shifts every later
ordinal and turns replay into a duplicate. An \emph{ancestry tag} survives relocation, but the
tag must be authored, and a call that moves into a new frame changes its parent. A
\emph{semantic role} survives relocation, but the role is a declaration outside the runtime's
verification, with the same status as a version marker. We choose the local, host-derived
identity because a pre-spend check can compute it without the model's cooperation, and we
state its cost. E7 measures the alternative, in which payload identity accepts every relocation and
returns a result committed at another callsite twice in four repairs
(Table~\ref{tab:e7}).

\begin{table}[t]
\centering
\footnotesize
\setlength{\tabcolsep}{2pt}
\caption{What each identity provides. \emph{Pre-spend} means the host can compute the
identity before a container starts without an authored marker. The last column is the
measured or constructed outcome. Callsite identity is the only row that is both
host-derived and free of wrong reuse, and its cost is the relocation stop.}
\label{tab:identity-choice}
\begin{tabular}{@{}>{\raggedright\arraybackslash}p{2.55cm}cccc@{}}
\toprule
Identity & Reloc. & Misbind. & Pre-spend & Outcome \\
\midrule
payload \((\mathrm{cap},H(p))\) & yes & no & yes & wrong reuse \\
arrival ordinal & no & yes & yes & duplicate \\
source line & no & yes & yes & stops \\
global ordinal & yes & no & no & duplicate \\
ancestry tag & yes & yes & no & authored \\
semantic role & yes & yes & no & declared \\
callsite \((\sigma,\pi,\omega)\) & no & yes & yes & pre-spend stop \\
\bottomrule
\end{tabular}
\end{table}

Identity and payload vary independently, which gives four cases. A call at a committed
identity with the same payload replays and pays nothing. A call at a committed identity with
a different payload raises \textsf{Divergence}. A call at a new identity with a payload
already paid for at the same site and path is a semantic repeat, which the ledger names. A
call at a new identity with a new payload is new work. The gate guarantees the first case before a container starts, the digest enforces the second at dispatch, and the ledger covers the third.

\subsection{Shifts that produce a new identity}
\label{sec:boundary}

The identity has three components, so a repair has three ways to produce a call that is new
but semantically old. Table~\ref{tab:shifts} lists them and names the mechanism that catches
each one.

\begin{table}[t]
\centering
\footnotesize
\caption{Where each shift is caught. The static gate compares sites, capabilities,
and call expressions, while \(\pi\) and \(\omega\) are runtime quantities.}
\label{tab:shifts}
\setlength{\tabcolsep}{2pt}
\begin{tabular}{@{}>{\raggedright\arraybackslash}p{1.55cm}>{\raggedright\arraybackslash}p{1.0cm}>{\raggedright\arraybackslash}p{1.45cm}>{\raggedright\arraybackslash}p{0.95cm}>{\raggedright\arraybackslash}p{2.45cm}@{}}
\toprule
Shift & Gate & Digest & Ledger & Consequence \\
\midrule
\(\omega\) grows, same payload & accepts & passes & blocks & the ledger names the repeat \\
\(\omega\) grows, new payload & accepts & passes & -- & new work (intended) \\
\(\pi\) shifts, other payload & accepts & \textsf{Divergence} & -- & stop, no duplicate \\
\(\pi\) shifts, same payload & accepts & hit & -- & interchangeable \\
duplicate insertion & accepts & hit & -- & inserted call replays, original is new \\
new callsite & -- & -- & -- & duplicate (residual case) \\
\bottomrule
\end{tabular}
\end{table}

\paragraph{Relocation}
A fourth shift changes the identity without changing the command sequence, because it moves a
committed call into a helper. The gate stops it (Proposition~\ref{prop:gate-incomplete}), and
Section~\ref{sec:baselines} measures the difference against a durable engine's reset, which
completes the same repair.

\paragraph{Occurrence extension}
A script that retries with a fixed payload, for example a loop over \texttt{range(3)} whose
body calls \texttt{image.generate} with a constant prompt, commits \(\omega=0,1,2\). Widening
the bound to 6 does not touch \(\sigma\), the capability, or the call expression, so the gate
accepts. Occurrences 3, 4, and 5 are fresh identities, so the runtime rule executes them. The
ledger names the repeat before the call is paid.

\paragraph{Path shift}
Inserting a fanout branch ahead of a committed one moves every later branch by one path index.
The gate still accepts, and the branch that now reaches a committed path carries a different
payload, so the runtime digest raises \textsf{Divergence} and the run stops. No call is
duplicated, but the counterexample is a digest mismatch rather than a structural explanation.
When the shifted payload is identical the two branches are interchangeable, and the run
reproduces the committed artifact.

Without the ledger, widening the bound spends three more calls, and under \(\mathsf{deny}\) it spends nothing, while a path shift stops after one call in either case. Together the three mechanisms give identity-level once. The ledger covers the semantic repeat for a stable
payload, and a prompt that embeds a timestamp or a random seed lies outside its scope by
construction.

\subsection{Crash handling}
\label{sec:crash}

Let \(k\) be the number of calls in flight when the process crashes, which is the in-flight
concurrency measured per crash.

\begin{proposition}[Policy-dependent crash window]
\label{prop:crash}
Assume the attempt record is persisted before dispatch.
\begin{enumerate}
\item Under fail-closed, an unresolved side-effecting attempt raises
\(\mathsf{AttemptUnresolved}\). The \(k\) identities produce no silent duplicate until
reconciliation, and the run waits for the operator's decision.
\item Under explicit abandon-and-retry, an operator may declare an attempt abandoned and
allow execution. If the provider completed and only the commit was lost, this can repeat up to \(k\) effects per crash, and the decision is operator-visible and auditable.
\item Under fail-open, a failed attempt write permits dispatch, and the bound becomes the
operator's responsibility.
\item A provider-side idempotency key or two-phase commit closes the window at the provider,
and the contract composes with it.
\end{enumerate}
\end{proposition}

The default policy is detect-and-fail-closed. An unresolved side-effecting call is
detectable, and it is not replayed by default. Bounded duplication, at most \(k\) per crash,
arises only after an explicit abandonment. A provider idempotency key composes with the
contract and closes the window at the provider, while the contract supplies identity-level
once, detection, and a bounded window.

The broker writes an attempt record before dispatch, so a crash leaves the attempt without a committed result, and on replay a side-effecting capability raises
\textsf{AttemptUnresolved}. The operator either writes back a known result or declares the
attempt abandoned. Abandonment permits a retry and can repeat an effect that the provider had
already completed, so the decision is recorded and auditable.

\subsection{Protocol walks}

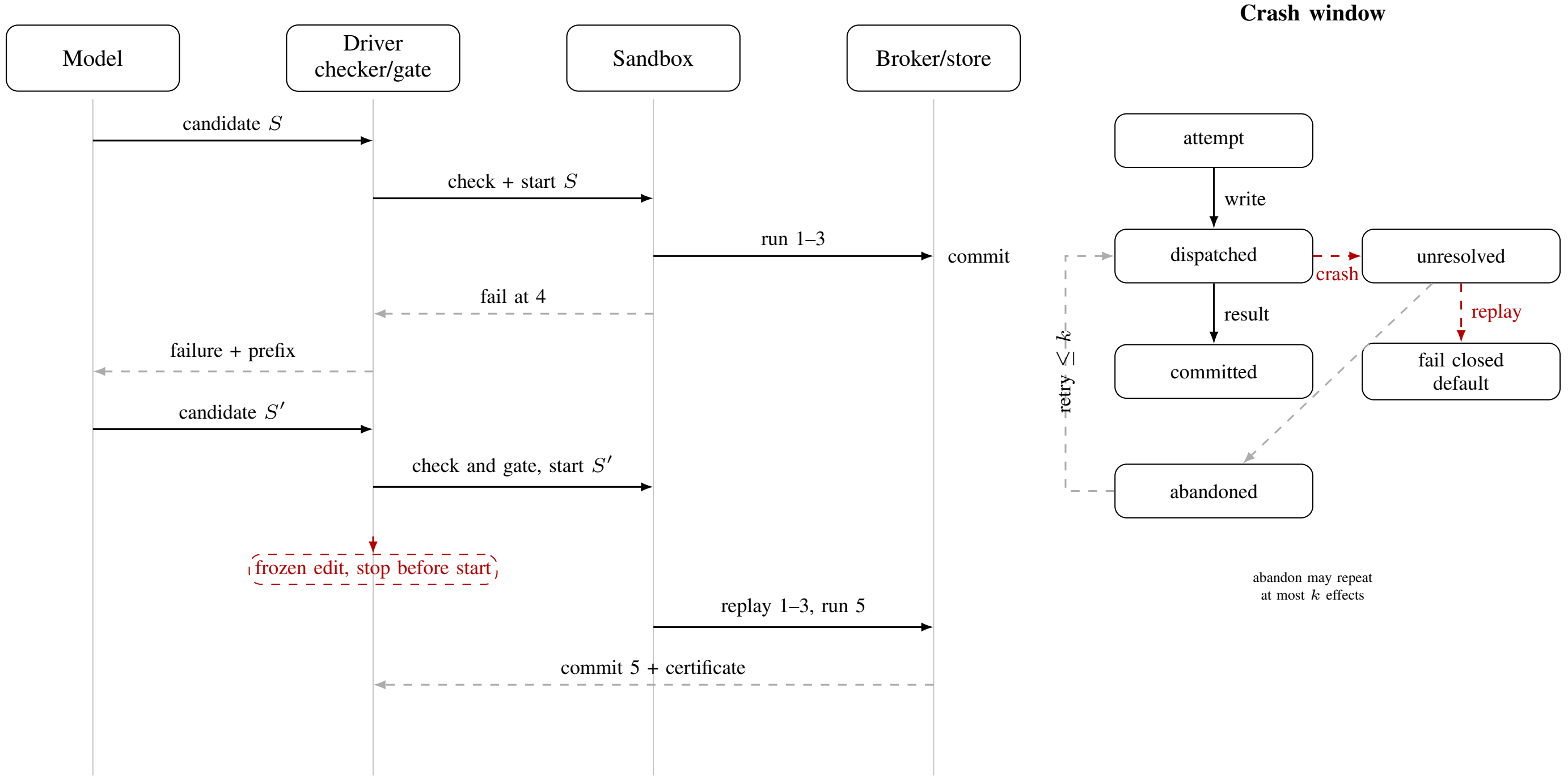
\begin{figure*}[t]
\centering
\resizebox{\textwidth}{!}{%
\begin{tikzpicture}[
  actor/.style={draw, rounded corners, align=center, font=\footnotesize,
                minimum width=2.1cm, minimum height=0.8cm},
  state/.style={draw, rounded corners, align=center, font=\scriptsize,
                minimum width=2.4cm, minimum height=0.65cm},
  msg/.style={-{Latex[length=1.5mm]}, semithick},
  ret/.style={-{Latex[length=1.5mm]}, semithick, dashed, draw=gray!65},
  redmsg/.style={-{Latex[length=1.5mm]}, semithick, dashed, red!70!black},
  lab/.style={font=\scriptsize, align=center},
  note/.style={draw, dashed, rounded corners, font=\scriptsize, align=center, inner sep=2pt}
]
\node[actor] (m) at (0,0) {Model};
\node[actor] (d) at (3.4,0) {Driver\\checker/gate};
\node[actor] (s) at (6.8,0) {Sandbox};
\node[actor] (b) at (10.2,0) {Broker/store};
\foreach \x in {0,3.4,6.8,10.2} \draw[gray!45] (\x,-0.5) -- (\x,-8.7);

\draw[msg] (0,-1.0) -- node[above,lab]{candidate $S$} (3.4,-1.0);
\draw[msg] (3.4,-1.7) -- node[above,lab]{check + start $S$} (6.8,-1.7);
\draw[msg] (6.8,-2.4) -- node[above,lab]{run 1--3} (10.2,-2.4);
\node[lab,right] at (10.25,-2.4) {commit};
\draw[ret] (6.8,-3.1) -- node[above,lab]{fail at 4} (3.4,-3.1);
\draw[ret] (3.4,-3.8) -- node[above,lab]{failure + prefix} (0,-3.8);
\draw[msg] (0,-4.5) -- node[above,lab]{candidate $S'$} (3.4,-4.5);
\draw[msg] (3.4,-5.2) -- node[above,lab]{check and gate, start $S'$} (6.8,-5.2);
\node[note, text=red!70!black, draw=red!70!black] (rej) at (3.4,-6.2)
  {frozen edit, stop before start};
\draw[redmsg] (3.4,-5.8) -- (rej.north);
\draw[msg] (6.8,-6.9) -- node[above,lab]{replay 1--3, run 5} (10.2,-6.9);
\draw[ret] (10.2,-7.6) -- node[above,lab]{commit 5 + certificate} (3.4,-7.6);

\begin{scope}[xshift=12.4cm]
\node[font=\footnotesize\bfseries] at (2.4,0.55) {Crash window};
\node[state] (att) at (1.2,-1.0) {attempt};
\node[state] (disp) at (1.2,-2.4) {dispatched};
\node[state] (com) at (1.2,-3.8) {committed};
\node[state] (unr) at (4.2,-2.4) {unresolved};
\node[state] (fail) at (4.2,-3.8) {fail closed\\\scriptsize default};
\node[state] (aba) at (1.2,-5.25) {abandoned};
\draw[msg] (att) -- node[right,lab]{write} (disp);
\draw[msg] (disp) -- node[right,lab]{result} (com);
\draw[redmsg] (disp) -- node[below,lab]{crash} (unr);
\draw[redmsg] (unr) -- node[right,lab]{replay} (fail);
\draw[ret] (unr) -- (aba);
\draw[ret] (aba.west) -- (-0.6,-5.25) -- (-0.6,-2.4) -- (disp.west);
\node[lab,rotate=90] at (-0.6,-3.8) {retry $\le k$};
\node[font=\tiny,align=center] at (2.4,-6.4) {abandon may repeat\\at most \(k\) effects};
\end{scope}
\end{tikzpicture}%
}
\caption{Protocol walks. (a) A normal run, a tail repair, and the gate's
pre-spend stop. (b) The crash window. The attempt record is written before dispatch, and an unresolved attempt fails closed unless an operator abandons it, which can repeat at most \(k\) effects.}
\label{fig:protocol}
\end{figure*}

Figure~\ref{fig:protocol}(a) walks a normal run and a tail repair, and part (b) shows the
crash window.

\paragraph{Normal run}
The driver parses the script and runs the checker. On acceptance the host injects a
\verb|_site| keyword into each capability call, starts the sandbox, and passes the annotated
source. The host annotator owns \verb|_site| and accepts an existing value only when it
equals the computed callsite. The container reports \verb|(site,path,occurrence)| with every
RPC, and each frame carries a per-run random nonce that only the wrapper knows, which binds
the call to the host's annotation. The broker computes a payload digest and a slot key,
executes the handler on a miss, commits the result and the slot, and returns.

\paragraph{Repair after failure}
The driver loads the prefix and asks the model for a candidate, including the frozen sites
and the structured failure. The candidate is checked and then gated. A frozen edit raises a counterexample, and the driver retries without starting a container. If the candidate passes,
replay serves committed identities from the store and executes only new identities. The host
compares the frozen identities with those invoked, and a non-empty difference is an orphan,
recorded without stopping the run.

\subsection{Trust boundary and failure policy}
\label{sec:system}

Storage errors are fail-closed by default, so a failed slot listing or idempotency read raises and an outage surfaces as a visible failure. The safety claim assumes a store that answers
(Assumption~\ref{a:durable}), and the default is the safety choice. Explicit opt-outs
(\texttt{DYNAMIC\_*\_FAIL\_CLOSED=0}) make the permissive policy an operator decision.

The host trusts the image digest, the broker, and the store. The container's authority is the host's annotation table. The host supplies \verb|_site|, and each RPC is bound to that annotation by a per-run nonce. The image has no network and a read-only root, and provider credentials stay in the host. The gate and the runtime check are the enforcement points, and neither depends on the
model's cooperation.

\section{Implementation}
\label{sec:impl}

The prototype runs the contract inside a networkless Docker sandbox that never sees a
provider credential. A host service accepts a model-authored script as a job, validates it,
runs it in the container, and routes capability calls back to the host. The repository
contains the checker, the site annotator, the prefix gate, the broker, the run store, the
repair loop, the sandbox controller, and the certificate and orphan endpoints. The
implementation is Python 3.11, and the run store is SQLite in WAL mode.

\paragraph{Checker}
The checker is an AST pass over the script with thirteen rules, R00 to R12. It returns
\verb|violations| and \verb|facts|, and \verb|valid| is determined by the error-severity
violations. The checker never executes the source, and the repair loop runs the script rules and
skips the envelope rules, so authorization stays in the host's grant set.

\paragraph{Structural sites}
\verb|assign_sites| assigns each capability call a site from the enclosing function, the
block path, and the call ordinal. The identifier does not depend on line numbers, comments,
formatting, or payload values. R10 reserves \verb|_site| for the host, and the checker
enforces it. \verb|annotate| injects \verb|_site| without changing line numbers and accepts
an existing value only when it equals the computed callsite. The host keeps a
site-to-fingerprint whitelist and admits an RPC only when its site is in the whitelist. Each
frame carries a per-run nonce, which binds the call to the host's annotation.

\paragraph{Prefix gate and certificate}
{\raggedright
The loader \verb|Prefix.from_store| reads slot records, deduplicates by
\verb|(site, path, occurrence)|, and raises \verb|PrefixConflictError| on conflicting
duplicates. The loader supports permissive reads, and the gate calls it with
\verb|strict=True|, the default, so enforcement fails closed on a failed listing. The gate
emits \verb|removed| or \verb|capability_changed|, and a changed call expression emits
\verb|call_changed|. A failure raises before the container is created. The certificate is a
hash chain over entries sorted by \verb|(site, path, occurrence)|.
\verb|persist_certificate| writes it, and \verb|certificate_for_run| recomputes
differences.\par}

\paragraph{Broker and idempotency}
{\raggedright
The broker registers the capabilities in the run's grant set. Four of the nine catalog
capabilities are production-bound, namely \verb|storyboard.plan|, \verb|image.generate|,
\verb|review.visual|, and \verb|text.generate|. The other five serve dry-run authoring. Each
handler is wrapped by \verb|_idempotent|. The wrapper computes a canonical payload digest, a
slot key from \verb|(site, path, occurrence)|, and a result key that also includes the
payload digest. On a slot hit it compares the capability and the digest, and a mismatch
raises \verb|ReplayDivergenceError|. On a result hit it returns the stored result without
calling the provider. On a miss it writes an attempt record, calls the handler, and commits
the result and the slot. A \verb|source_digest| check backs up the gate's call-expression
check. A known handler exception is recorded as \verb|failed| and may be retried, while a
crash leaves \verb|attempt| for the crash policy.\par}

\paragraph{Sandbox}
\verb|SandboxRunner| launches the fixed runtime image with no network, a read-only root
filesystem, a non-root UID, dropped capabilities, \verb|no-new-privileges|, a seccomp
profile, cgroup limits on CPU, memory, and processes, and tmpfs workspaces. The source,
inputs, and output are size-capped, and capability calls use a bounded file RPC rather than
a network socket. The controller validates the image digest, the resource controller, and the
isolation profile before execution. An acceptance script covers sixteen checks, including
network isolation, read-only filesystem, non-root execution, seccomp, RPC authorization,
output limits, timeout, artifact export, budget downgrade, concurrency, and capability audit.

\paragraph{Repair loop}
\verb|repair_dynamic_source| performs at most \verb|max_attempts| rewrites. It validates the
candidate with the checker and the prefix, and it additionally requires that the number of
capability calls does not decrease, which catches a truncated function that still parses. The
prompt contains the frozen sites, their source lines, the original committed script, the last
failure, and the structured counterexamples. The loop returns the candidate and the
per-attempt feedback, and execution is a separate caller decision.

\paragraph{Audit surfaces}
The host records prefix verification, sandbox start and exit, capability start, denial, and
completion, orphan reports, and certificates. The certificate and orphan endpoints expose the
recomputed summary and the frozen-but-not-invoked identities. The certificate is a digest
manifest that verification recomputes from the store. Provider credentials and the token
budget remain in the host process.

\paragraph{Reproducibility}
The checker, the annotator, the prefix gate, the broker, the run store, the repair loop, the
E1 to E4 scripts, and the P1-C and P2 mutation harnesses are in the authors' repository. E1
uses a stored task set and stored model outputs, P1-C uses stored outputs, and P2 uses stored
mutants. The E2b, E4, and P2b commands run without Docker, network, provider credentials, or
an API key. A reproducibility artifact accompanies the submission and carries the
pre-registration specifications, the frozen task sets, the exact commands, the raw scores,
and an offline smoke script. The E1 task set is frozen at 60 tasks, and the artifact records
its definition, the pinned container image, the P1-C task and manifest identifiers, the P2
and P2b base identifiers, and the per-bucket cost records. Model IDs are the
provider-served IDs on the run date.

\section{Evaluation}
\label{sec:eval}

The evaluation reports spend, identity choice, repair quality, portability, and the cost of
authoring and checking. A \emph{paid call} is a dispatch attempt, including one that the
provider rejects, and a \emph{successful execution} returns an artifact. Every table that counts
executions names which of the two it counts.

\subsection{Research questions and study design}

Six questions organize the study, and the next paragraph maps them to the studies.

RQ1. What does a repair spend, and how does a durable engine compare?

RQ2. Which identity should the contract use, and what does the writable set admit?

RQ3. Can the model produce repairs that the gate admits?

RQ4. Does the contract transfer to another domain?

RQ5. What do authoring and checking cost?

RQ6. What do reliability and overhead cost?

All studies share one host, one run store, and one sandbox image, and every spend study uses
a single run id per episode. RQ1 compares a contract run with a full rerun and with a durable
engine's reset. E2b replays real model repairs at the replay layer, E3 runs the protocol
inside Docker, E5 and E5b bill a live provider, and the engine study ports five cases to
Temporal. RQ2 re-runs the E2b cases under four identity functions that share the slot,
divergence, and commit code and differ only in the key, and it applies seven deterministic
edits to real prefixes. RQ3 measures repair on eight real failures and on a 32-script
benchmark drawn from model output. RQ4 applies the same identity and replay checks to twelve
customer-support tasks. RQ5 covers authoring on 60 tasks, execution on the accepted scripts, held-out transfer, task success with human scoring, and checker validity on two mutation corpora. RQ6 covers the crash window, store failures, orphan reporting, and the latency of the gate.

Four quantities are counted. A \emph{committed identity} is an identity with a stored record, and a \emph{replay} serves one from the store. A \emph{duplicate}
executes a call whose work is already committed under the same run. \emph{Wrong reuse} serves a record that belongs to a different callsite, and only E7 reports it.

Two baselines bound the spend. A \emph{full rerun} executes the repaired script from the
start with an empty store. A \emph{reset} asks a durable engine to restart from the last
workflow task before the failed activity, which keeps the activities completed before that
point. E7 adds three identity functions behind the same slot and commit layer, and two
constructed probes test a payload cache.

Most studies run offline, because E2b, E1, E1b, P1-C, P2, P2b, and E7 use stored inputs and stored model outputs and need neither Docker nor a provider key. E3 runs on Docker 28.0.1 for
linux/arm64 with the pinned image, and E4 uses a real child process and the production SQLite
store. E5 and E5b bill two image providers, and N1 bills a text provider.

\subsection{Repair cost and engine comparison (RQ1)}
\label{sec:spend}

RQ1 asks what a repair spends, and we measure it at three levels. E2b replays real model repairs
against a local harness, E3 runs the protocol inside Docker, and E5 and E5b bill a live
provider. The subsection ends with a durable engine on the same repairs and with the cost
model that turns call counts into money and time.

\paragraph{Replay of real model repairs (E2b)}

E2b executes the original script against a deterministic local provider until the \(k\)-th
capability execution fails, persists the committed slots, and then runs the repaired script on
the same run id. The harness uses the production identity, slot, idempotency, gate, and
certificate code, so replay is isolated from container and provider variance. The repaired
source is the output that E2 produced for the same failure. Four failure points come from three scripts. One loop case is run at two failure points, the second iteration's image and review, labeled \texttt{5698a} and \texttt{5698b}.

\begin{table}[t]
\centering
\small
\setlength{\tabcolsep}{4pt}
\caption{E2b, injected tail failures on real model repairs. Cmt.=committed
identities, Rep.=replayed, New=newly executed. Tot. and Naive count successful
executions over the failed run plus the repair. The injected failure is a paid attempt but
not a success. Naive executes the repair without the gate. Cmt. sums to 11, with eight
in the three admitted runs and three in the stopped run. \(\dagger\) marks a gate stop
before spending, with the committed site removed.}
\label{tab:e2b}
\begin{tabular}{@{}lrrrrrr@{}}
\toprule
Case & \(k\) & Cmt. & Rep. & New & Tot. & Naive \\
\midrule
\texttt{e388} & 2 & 3 & 3 & 1 & 4 & 7 \\
\texttt{a425}\(^{\dagger}\) & 2 & 3 & 0 & 0 & 3 & 7 \\
\texttt{5698a} & 3 & 2 & 2 & 2 & 4 & 6 \\
\texttt{5698b} & 4 & 3 & 3 & 1 & 4 & 7 \\
\midrule
Total & -- & 11 & 8 & 4 & 15 & 27 \\
\bottomrule
\end{tabular}
\end{table}

Across the three admitted repairs, eight committed identities are replayed and none is
executed again. Across all four runs, no committed identity is re-executed, no orphan is
reported, and every certificate recomputes with \texttt{differences=[]}. The three admitted
repairs make 12 successful executions where a full rerun makes 20. Counting the committed
prefix of the stopped run, the episode is 15 against 27, and the four injected failures
account for the difference between 15 and the 19 dispatch attempts. The parallel case commits
three of four branches and replays all three, and the loop case replays occurrence 0 and
executes occurrence 1, so the result holds for parallel and for sequential code.

The \texttt{a425} case shows the boundary. Its repair passes R00 to R12 and preserves the capability multiset, but it moves the fanout call into a helper. The committed site
\texttt{orchestrate/\#1} disappears, and the gate stops the candidate before spending, with
the status \texttt{removed}. Under structural identity the moved call is new and would
execute, which is the trade-off the contract accepts. The shipped repair prompt lists the
frozen sites, and RQ3 measures what that listing changes.

\paragraph{End-to-end in Docker (E3)}

E3 runs the pinned image under Docker 28.0.1. One script is submitted three times under the
same run id and store. The first submission, v1, calls \texttt{echo(n=1)} and
\texttt{echo(n=2)} and then raises. The second, v2, appends \texttt{echo(n=3)} and changes
nothing else. The third, v3, edits the committed \texttt{n=1} payload to \texttt{n=9}.
\texttt{echo} is a local counting handler that appends its argument to a list on every real
execution, and the production idempotent layer wraps it with structural slots.

v1 executes two calls and fails. v2 replays both and executes only the third, so the
cumulative handler log is $[1,2,3]$ rather than a rerun of all three. A control that runs v2
without the idempotent layer logs $[1,2,3]$ again, which locates the saving in replay rather
than in the handler. v3 is stopped before it starts. The gate raises
\texttt{dynamic\_prefix\_changed} with the counterexamples \texttt{call\_changed} and
\texttt{removed}, and the handler log stays at three. The certificate recomputed from the
slots has \texttt{differences=[]}, and the experiment reports 8 of 8 checks. The episode
costs three real executions instead of five.

\paragraph{A live provider (E5, E5b)}

E5 puts the spend on a paid image API. A four-shot script requests a 1K image per shot, and
the fourth call uses an unsupported size, so the provider returns
\texttt{400 InvalidParameter} after three committed images. The repair loop rewrites the
script, the gate accepts the candidate, and the run replays on the same run id and store.

\begin{table}[t]
\centering
\small
\caption{E5 real-provider repair (Seedream, 1K, one host). \emph{Calls} are HTTP
requests to the provider. The naive row is a full rerun of the repaired script,
measured over three control runs on 2026-10-31 (83--89\,s per control run).}
\label{tab:e5}
\begin{tabular}{@{}lrrr@{}}
\toprule
 & Calls & Images & Wall clock \\
\midrule
First run (3 ok, 1 rejected) & 4 & 3 & 109.1\,s \\
Repair (one model attempt) & 0 & -- & -- \\
Replay (same run id) & \textbf{1} & 1 & 35.6\,s \\
\midrule
Contract path, total & 5 & 4 & 144.7\,s \\
Naive full rerun (measured, 3 runs) & 8 & 7 & -- \\
\bottomrule
\end{tabular}
\end{table}

Three committed identities are replayed with zero provider calls, and the replayed URLs are
byte-identical to those of the first run, so the result is a stored artifact rather than a
recomputation. The contract path makes five calls and produces four images in 144.7\,s, while
a full rerun makes eight calls and seven images. The model's edit replaces a data-dependent
size expression with a valid one. The failing identity never committed, so the change falls
outside the frozen prefix.

E5b adds four runtime failure classes on a second live provider, \texttt{gpt-image-2} through
a LiteLLM proxy. The failures are an invalid \texttt{n} (404), an invalid model (403), a
client timeout (0.01\,s), and a three-way fanout with one invalid branch. In each case the
original run commits one or two calls and fails at the next, the repair changes a variable
outside the call expression, the gate accepts, the committed prefix replays, and the failed
identity executes once. Across the failed run and the repair the contract makes 3 calls where
a full rerun makes 4, and on the partial batch it makes 4 against 6. Replayed artifacts are
byte-identical and every certificate has no differences. The four classes and their controls
cost 22 provider calls and \$0.71. The timeout is the bounded window of
Proposition~\ref{prop:crash}. The client aborted before the provider replied, so the provider
may have completed the request, and the contract re-executes the failed identity and records
the window.

\paragraph{A durable-execution engine (G1)}
\label{sec:baselines}

We ported five cases to Temporal (CLI 1.9.1, server 1.32.0, Python SDK 1.34.0) and ran three
recovery paths on the same committed history. The paths are a new run, a reset to the last
workflow task before the failed activity, and a replay of the repaired source against the
original history. Temporal activities are deterministic local stand-ins that log every
dispatch. Table~\ref{tab:g1} reports the episode totals, which count the failed run and the
recovery run together.

\begin{table*}[t]
\centering
\small
\setlength{\tabcolsep}{4pt}
\caption{The same repairs under the contract and under Temporal. Each contract/reset/naive
cell is attempts/successful executions for the whole episode. Reset is Temporal's standard
recovery, to the last workflow task before the failed activity, and naive is a new run of the
repaired script. \texttt{e388}, \texttt{5698}, \texttt{e5}, and \texttt{a425} are real
repairs. \texttt{s1} is a constructed probe that isolates the committed-payload case, and its
reset returns a stale artifact, because the first shot is the recorded result for the old
prompt. $^{\dagger}$Replay accepts but returns the recorded failure, and $^{\ddagger}$replay
accepts and returns the recorded result.}
\label{tab:g1}
\begin{tabular}{@{}p{2.0cm}p{3.0cm}cccp{1.25cm}>{\raggedright\arraybackslash}p{4.45cm}@{}}
\toprule
Case & Shape & Contract & Reset & Naive & Replay & Outcome \\
\midrule
\texttt{e388} & parallel fanout & \textbf{5/4} & 7/6 & 8/7 & ok & contract replays committed branches \\
\texttt{5698} & sequential loop & 5/4 & 5/4 & 7/6 & ok & equal spend, four artifacts each \\
\texttt{e5} & sequential, 400 & 5/4 & 5/4 & 8/7 & ok$^{\dagger}$ & equal spend, reset applies the new payload, replay does not \\
\texttt{a425} & relocate into helper & 4/3, no output & 7/6 & 8/7 & nondet. & gate stops before spending, engine resets through \\
\texttt{s1} & committed payload & 4/3, no output & 5/4, stale & 8/7 & ok$^{\ddagger}$ & digest stops, engine reuses the old result \\
\bottomrule
\end{tabular}
\end{table*}

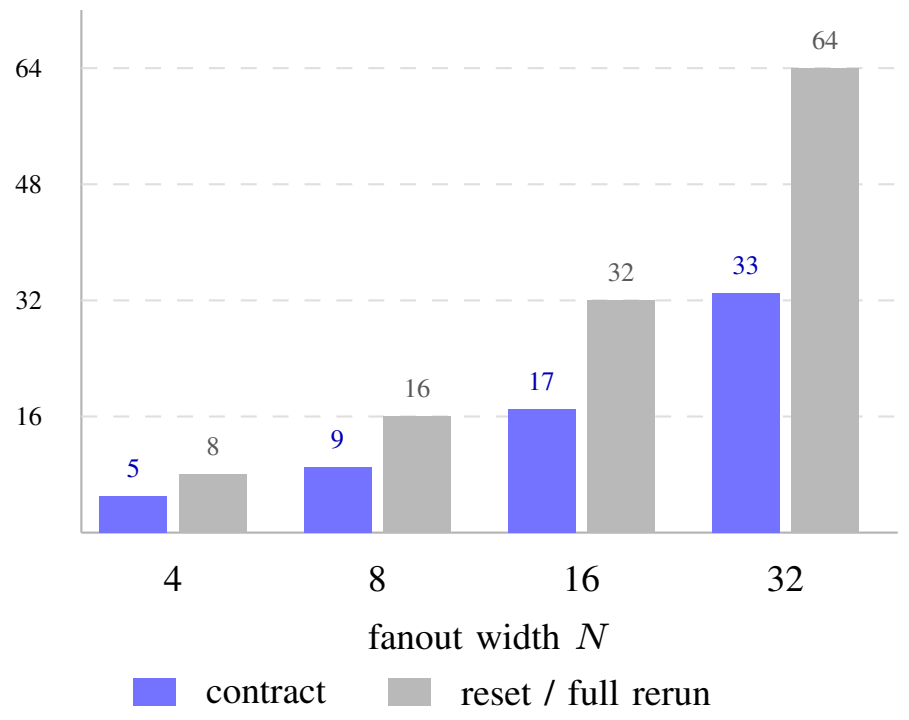
\begin{figure}[t]
\centering
\begin{tikzpicture}[x=1.35cm,y=0.048cm,font=\scriptsize]
  \draw[gray!60] (0.55,0) -- (4.55,0);
  \draw[gray!60] (0.55,0) -- (0.55,72);
  \foreach \v in {16,32,48,64} {
    \draw[gray!25,dashed] (0.55,\v) -- (4.55,\v);
    \node[left,font=\tiny] at (0.45,\v) {\v};
  }
  \foreach \i/\n/\c/\r in {1/4/5/8,2/8/9/16,3/16/17/32,4/32/33/64} {
    \fill[blue!55] (\i-0.36,0) rectangle (\i-0.03,\c);
    \fill[gray!55] (\i+0.03,0) rectangle (\i+0.36,\r);
    \node[below] at (\i,-2.0) {\n};
    \node[above,font=\tiny,blue!70!black] at (\i-0.195,\c) {\c};
    \node[above,font=\tiny,gray!70!black] at (\i+0.195,\r) {\r};
  }
  \node[below] at (2.55,-10.0) {fanout width $N$};
  \node[fill=blue!55,minimum width=0.28cm,minimum height=0.18cm,inner sep=0,
        anchor=west] (l1) at (0.8,-22.0) {};
  \node[right=2pt of l1,anchor=west] (t1) {contract};
  \node[fill=gray!55,minimum width=0.28cm,minimum height=0.18cm,inner sep=0,
        anchor=west,right=8pt of t1] (l2) {};
  \node[right=2pt of l2,anchor=west] {reset / full rerun};
\end{tikzpicture}
\caption{Per-identity replay against workflow-task reset on an $N$-way fanout with one
failing branch. The contract makes $N+1$ calls, and reset and a full rerun make $2N$ when
the batch is restarted. Measured
at $N=4,8,16,32$, with three repeats each. The gap is $N-1$.}
\label{fig:fanout}
\end{figure}

On sequential tail repairs the reset matches the contract. For \texttt{5698} and \texttt{e5}
both approaches make five attempts and produce four artifacts. Reset keeps the activities
completed before the failed one and re-executes from there, which costs the same as replaying
committed identities. The advantage appears when the failure falls inside a fanout. On
\texttt{e388} the reset re-executes the three branches after its reset point and makes seven
attempts, while the contract replays three committed branches and executes one, for five. The gap grows with width. With the reset point fixed before the batch, the contract makes \(N+1\)
attempts and the reset makes \(2N\), and the measured widths \(N=4,8,16,32\) save 5 against
8, 9 against 16, 17 against 32, and 33 against 64 calls, with three repeats each.
Figure~\ref{fig:fanout} isolates that shape.

The identity function also differs. Moving a committed call into a helper leaves
the command sequence unchanged, so Temporal resets through the \texttt{a425} repair and
completes it in seven attempts, while our gate stops the candidate before spending. The stop
is correct under structural identity, because the moved call has a new \(\sigma\) and would
execute. A committed-payload change lies outside the view of replay. On \texttt{s1}, a
constructed probe, Temporal returns the recorded result for the old prompt, with five
attempts and a stale first shot, while the digest raises \textsf{Divergence} before dispatch,
with four attempts and no output.

Replay itself differs as well, because it preserves the recorded outcome. On \texttt{e5} the
repaired fourth call has a new payload, but the history records a 400 at that position.
Replay returns the recorded failure, and the new payload needs a reset, a new run, or a
versioned branch. The reset column is therefore the fair comparison, and the replay column is
not.

On a second provider, \texttt{gpt-image-2} through a LiteLLM proxy, a six-way fanout with one
empty-prompt failure makes seven paid calls and returns six images, five of them replayed
byte-identically, while a full rerun makes twelve. The experiment cost \$0.47 for twelve
billed images across the contract run and its control, and the contract path alone is seven
calls, six images, and \$0.23.

We also asked the model to make the payload edit. Twenty scripts with a committed prefix each
carry a constant that feeds the first committed call, and the repair instruction changes the
brand color and asks for the constant to be updated without touching the call expression. All
twenty candidates keep the call expression and pass the gate. At dispatch the digest raises
\textsf{Divergence} in all twenty, while a replay-only engine returns the recorded artifact in
all twenty. A payload cache completes all twenty with 48 executions, the supersede append
completes all twenty at the same count, and a full rerun executes 68. These are instructed
edits, so they show that the model can produce the case. They do not measure how often it
would do so unprompted.

A reset replays recorded results and re-executes from the reset point. The operator chooses that point, and replay is payload-blind. A repair that changes the payload of a committed call is therefore either replayed with the old result or, when the reset point moves before that activity, re-executed
and paid again. The contract closes both gaps with a source-level set of allowed edits that is checked before a container starts, a per-identity replay, and a payload digest.

DBOS 3.3 with step-level fork reproduces the same limit. On a three-step sequential workflow it
re-executes only the failed step, for four calls and equal spend. On a four-way fanout it
re-executes three steps, for seven calls against five for the contract. Step-level fork is
finer than Temporal's workflow task, but a completed branch after the fork point is still
re-executed, so the fanout gap follows from the granularity of the identity.

A repair that must change a committed payload can append a new call at a new identity and use
its result. The committed call replays and its artifact is left out of the product. On
\texttt{s1} this costs two new calls, the failed fourth and the replacement, and returns the
corrected artifact, while a full rerun of the repaired script costs five new calls and re-pays
the three committed images. The gate admits the append because it adds a callsite, and the
digest stays quiet because the new call is a new identity.

\paragraph{Workload and cost model}

The platform history contains 52 unique scripts, and 21 of them contain no capability call. Fourteen use \texttt{parallel}, and 31 use a loop.
Thirty-two fail the current checker, and of those 22 are old-API migrations, eight are the E2 failures, and two are out of scope, one with an R10 import and one with a non-production capability. The 30 recorded jobs are development runs, and the store holds 165 slot records,
so the cost evidence is unit costs plus the measured cost of real runs.

Prices vary by provider, model, resolution, duration, and plan, so we carry the per-call price
\(p\) as a variable and use the catalog estimate of \$0.02 per image call as one calibration
point. Every derived figure is linear in \(p\), and a repair that replays \(k\) calls saves
\(k\cdot p\). At \(p=\$0.02\) a repair that re-executes three images costs \$0.06, and a
fanout saves seven calls, or \$0.14, at \(N=8\), and 31 calls, or \$0.62, at \(N=32\). The catalog prices a video call at \$0.50 and a video edit at \$0.60. A six-shot repair that replays three shots saves \(3p\), which is \$1.50 at \(p=\$0.50\), and an \(N\)-way video
fanout saves \((N-1)\cdot p\), which is \$2.50 at \(N=6\).

\paragraph{Net benefit}

The repair pays off in both time and money. It saves \(N-1\) provider calls on an \(N\)-way
partial fanout and adds at most 1.19 rewrite attempts, measured as 2.19 against 1.00 in
the controlled ablation of RQ3, while the production path of E2 averages 1.12. The measured latencies are 17.2\,s per rewrite (154.4\,s over nine E2 attempts) and 27.3\,s per image call (109.1\,s for four calls in E5). At those latencies the net time advantage is 61\,s and three calls at \(N=4\),
which is 82\,s of provider time against 20.5\,s of rewriting, and 826\,s and 31 calls at
\(N=32\). The break-even point for a rewrite is \((N-1)\cdot p/1.19\), which is \$0.05 at
\(N=4\) and \$0.52 at \(N=32\) for images, and \$1.26 and \$13.03 for video. A deployment can
apply the rule to its own prices.

\subsection{Identity and the writable set (RQ2)}
\label{sec:identity}

RQ2 asks which identity the contract should use and what the choice costs. E7 turns the
comparison into a measurement on the same four repairs, and the acceptance study measures
what the gate admits on real prefixes.

\paragraph{Four identity functions (E7)}
\label{sec:e7}

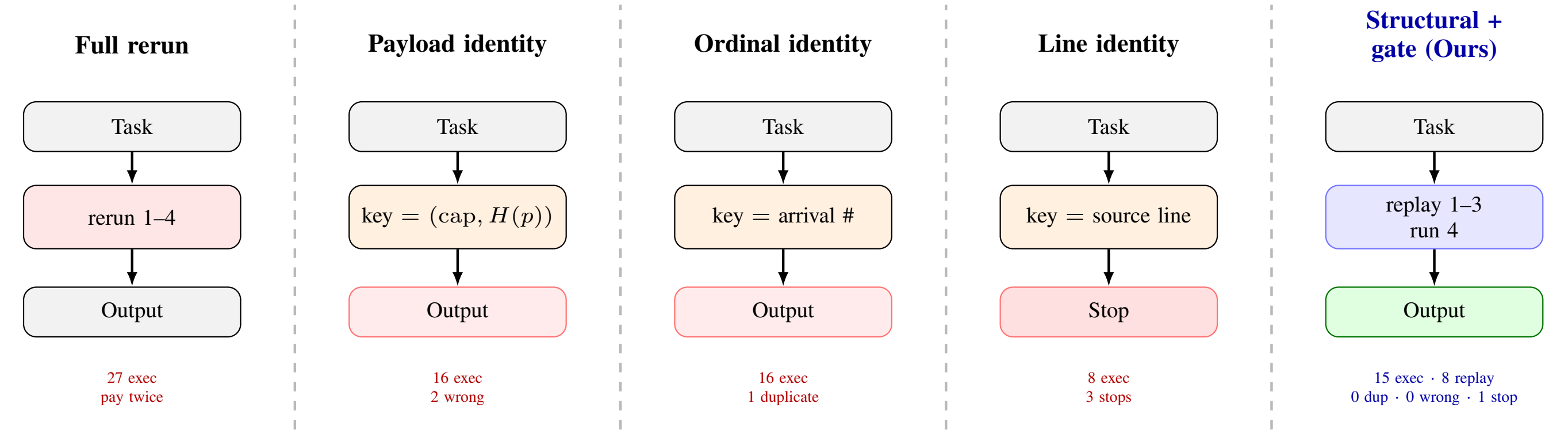
\begin{figure*}[t]
\centering
\resizebox{\textwidth}{!}{%
\begin{tikzpicture}[
  x=1cm, y=1cm,
  box/.style={draw, rounded corners, align=center, font=\scriptsize,
              minimum width=2.4cm, minimum height=0.55cm, inner sep=2pt},
  strat/.style={box, text width=2.25cm, minimum height=0.7cm},
  title/.style={font=\footnotesize\bfseries, align=center, text width=2.8cm},
  badge/.style={font=\tiny, align=center, text width=2.7cm},
  arr/.style={-{Latex[length=1.6mm]}, thick},
  sep/.style={dashed, gray!55, thick}
]
\node[title] at (1.5,2.25) {Full rerun};
\node[box, fill=gray!10] (t1) at (1.5,1.35) {Task};
\node[strat, fill=red!10] (s1) at (1.5,0.35) {rerun 1--4};
\node[box, fill=gray!10] (o1) at (1.5,-0.7) {Output};
\draw[arr] (t1) -- (s1);
\draw[arr] (s1) -- (o1);
\node[badge, red!70!black] at (1.5,-1.55) {27 exec\\pay twice};
\node[title] at (5.1,2.25) {Payload identity};
\node[box, fill=gray!10] (t2) at (5.1,1.35) {Task};
\node[strat, fill=orange!12] (s2) at (5.1,0.35) {key $=(\mathrm{cap},H(p))$};
\node[box, fill=red!8, draw=red!55] (o2) at (5.1,-0.7) {Output};
\draw[arr] (t2) -- (s2);
\draw[arr] (s2) -- (o2);
\node[badge, red!70!black] at (5.1,-1.55) {16 exec\\2 wrong};
\node[title] at (8.7,2.25) {Ordinal identity};
\node[box, fill=gray!10] (t3) at (8.7,1.35) {Task};
\node[strat, fill=orange!12] (s3) at (8.7,0.35) {key $=$ arrival \#};
\node[box, fill=red!8, draw=red!55] (o3) at (8.7,-0.7) {Output};
\draw[arr] (t3) -- (s3);
\draw[arr] (s3) -- (o3);
\node[badge, red!70!black] at (8.7,-1.55) {16 exec\\1 duplicate};
\node[title] at (12.3,2.25) {Line identity};
\node[box, fill=gray!10] (t4) at (12.3,1.35) {Task};
\node[strat, fill=orange!12] (s4) at (12.3,0.35) {key $=$ source line};
\node[box, fill=red!12, draw=red!55] (o4) at (12.3,-0.7) {Stop};
\draw[arr] (t4) -- (s4);
\draw[arr] (s4) -- (o4);
\node[badge, red!70!black] at (12.3,-1.55) {8 exec\\3 stops};
\node[title, blue!65!black] at (15.9,2.35) {Structural +\\gate \textbf{(Ours)}};
\node[box, fill=gray!10] (t5) at (15.9,1.35) {Task};
\node[strat, fill=blue!10, draw=blue!55] (s5) at (15.9,0.35) {replay 1--3\\run 4};
\node[box, fill=green!12, draw=green!45!black] (o5) at (15.9,-0.7) {Output};
\draw[arr] (t5) -- (s5);
\draw[arr] (s5) -- (o5);
\node[badge, blue!65!black] at (15.9,-1.55) {15 exec $\cdot$ 8 replay\\0 dup $\cdot$ 0 wrong $\cdot$ 1 stop};
\foreach \x in {3.3,6.9,10.5,14.1}
  \draw[sep] (\x,-2.0) -- (\x,2.7);
\end{tikzpicture}%
}
\caption{Five identity strategies on the same four E2b repairs. Each panel applies one
key function to the committed prefix and the failed tail, and the badges summarize the
outcome from Table~\ref{tab:e7}. Full rerun pays all calls again. Payload identity returns
two results recorded at a different callsite, ordinal identity repeats one committed call,
and line identity stops on three runs. Structural identity with the gate replays eight
committed identities and stops the one relocation. Solid arrows execute, dashed rules
separate strategies, and blue marks the contract.}
\label{fig:e7paradigms}
\end{figure*}

Related work positions step identity by argument, and E7 makes it a measurement. We re-ran the
four E2b cases under four identity functions that share the slot, divergence, and commit logic
and differ only in the key. The keys are the structural identity with the gate, the payload
pair of capability and digest, the arrival ordinal, and the source line.
Figure~\ref{fig:e7paradigms} draws the five strategies on one timeline, including the full
rerun, and Table~\ref{tab:e7} gives the counts.

\begin{table}[t]
\centering
\small
\caption{E7 identity baselines over the four E2b runs. \emph{Dup.} counts executions of
work already committed under the run, \emph{wrong} counts replay hits whose committed
record belonged to a different callsite, and \emph{stops} counts runs that end before
completion. The line row's execution count reflects early stops.}
\label{tab:e7}
\setlength{\tabcolsep}{3pt}
\begin{tabular}{@{}lrrrrr@{}}
\toprule
Scheme & Exec. & Replayed & Dup. & Wrong & Stops \\
\midrule
structural + gate & \textbf{15} & 8 & \textbf{0} & \textbf{0} & 1 \\
payload \((\text{cap}, H(p))\) & 16 & 11 & 0 & \textbf{2} & 0 \\
ordinal & 16 & 7 & \textbf{1} & 0 & 1 \\
line & 8 & 6 & 0 & 0 & \textbf{3} \\
\bottomrule
\end{tabular}
\end{table}

The structural column reproduces E2b exactly, with 15 executions against 27 and eight replays
from the store, which validates the harness. Payload identity completes every run, and two of
its hits land on a record committed by a different callsite, which is the wrong reuse that the
contract's identity prevents. Ordinal identity re-executes one committed call because arrival
order is not semantic order. Line identity stops on three of four runs, because fanout
branches and loop bodies share a line, so its execution count reflects the early stops.

The comparison runs four real repairs under one slot and commit layer, so the identity function is the only variable.

\paragraph{A payload cache}

E7 counts wrong reuse, and a constructed probe shows what it produces. We built 30
tasks with two identical-payload calls each. Fifteen require two distinct artifacts with
different roles, and 15 need one artifact because the second call is a redundant retry. Under
a payload cache the second call returns the first artifact, and under structural identity it
executes. On the different-role tasks the cache passes 0 of 15 and structural identity passes
15 of 15. On the same-role tasks both pass, and the cache uses 15 executions against 30 for
structural identity. A payload key is a deduplication key, so two roles with the same payload
share one key.

\paragraph{Shifts under the production gate}

We ran the two runtime shifts against the production gate, runtime rule, and ledger. Widening
a fixed-payload retry loop from three to six iterations is accepted by the gate and executes
three more calls. The policy \texttt{warn} records three semantic repeats without blocking,
\texttt{deny} raises before dispatch and spends nothing, and the default \texttt{auto} behaves
as \texttt{deny} here because \texttt{image.generate} has side effects. Inserting a fanout branch shifts every later path index. The runtime digest raises \textsf{Divergence} after one execution, so the addition reports as a failure, and all certificates recompute consistently.

\paragraph{What the gate admits}
\label{sec:freeze}

The writable set of the contract is callsite-level, and the implementation checks three things. A frozen callsite must exist, it must name the same capability, and its call-expression fingerprint must be unchanged. Statements around the call stay writable. We measure what that admits on
the four E2b failure points, first by edit class and then at scale.

We apply seven deterministic edits to the real prefix and ask the production gate.
Table~\ref{tab:admissible} shows the result, and the split is exact. The gate stops every edit to
the argument AST of a frozen call, whether it changes a literal or wraps a computed value. It
admits every edit to surrounding statements, an insertion of a non-call statement before the
call, and an append. It stops every edit that removes the callsite or relocates it into a new
block, and it reports \texttt{removed}.

\begin{table}[t]
\centering
\small
\setlength{\tabcolsep}{2pt}
\caption{What the gate admits, on three real scripts. \textbf{yes} = accepted,
\texttt{cc} = \texttt{call\_changed}, \texttt{rm} = \texttt{removed}, --- = not
applicable to that script.}
\label{tab:admissible}
\begin{tabular}{@{}>{\raggedright\arraybackslash}p{2.55cm}ccp{1.25cm}@{}}
\toprule
Edit & \texttt{e388} & \texttt{a425} & \texttt{5698} \\
\midrule
literal in payload & cc & --- & cc \\
computed payload & cc & --- & cc \\
statement outside & \textbf{yes} & --- & \textbf{yes} \\
insert a non-call statement & \textbf{yes} & \textbf{yes} & \textbf{yes} \\
wrap in new block & rm & rm & rm \\
delete a call & rm & --- & rm \\
append a new call & \textbf{yes} & \textbf{yes} & \textbf{yes} \\
\bottomrule
\end{tabular}
\end{table}

The split replicates beyond the three E2 scripts. We applied the same seven edits to twenty
E1 scripts that execute end to end against a deterministic local provider, each with its
maximal prefix, so every capability call is committed. Of 140 edit-script pairs, 136 apply,
and 60 are accepted against 76 stops. The stops are the predicted ones. Thirty-six are \texttt{call\_changed}, a literal or a computed value in the payload, and 40 are \texttt{removed}, a block wrap or a deletion. The three accepting classes accept on every
script, so the edit classes are a property of the gate rather than of the scripts.

Payload values inside a frozen call are frozen with it. The fingerprint hashes the argument
AST, so a literal change is as visible as an expression change. The eight real repairs in E2 edited the code around the call, for example a slice on a result list or a slice on the list joined into a prompt. The gate leaves that code writable, and the runtime digest checks it for committed identities.

The unit of the contract is the capability call, not the block and not the line, and the
choice has a cost. Relocating a committed call into a helper is routine, and the gate stops
it because the moved call is a new identity and would execute. The stop prevents a second
payment (Figure~\ref{fig:a425}), and Section~\ref{sec:baselines} measures it next to a reset
that completes the same repair. A finer freeze at committed occurrences would widen the
loop-repair space, while relocation would remain a trace-level check.

\subsection{Repair quality (RQ3)}
\label{sec:repair}

RQ3 asks whether the model can produce candidates that the gate admits, and how many rewrites
that takes. We measure the repair loop on eight real failures, on a 32-script benchmark, on a
corpus that covers four rules, and under a controlled prompt ablation.

\paragraph{Eight real failures (E2)}
\label{sec:e2}

E2 uses eight real model-authored scripts with contract violations and initial errors in R03
or R12. They are the strict subset of 32 checker failures from 52 historical scripts that have
replayable source and envelope, are deduplicated by digest, and stay within the four
production-bound capabilities. Twenty-two of the 32 are old-API R02 migrations and are
excluded. The repair driver validates each candidate with the checker and a non-decreasing
call-count guard, feeds structured counterexamples back to the model, and stops after three
rewrites. The model is \texttt{gemini-3.5-flash} at \texttt{temperature=0}.

\begin{table}[t]
\centering
\small
\caption{E2 static repair outcomes on eight real contract failures. All eight are
accepted by the checker. E2b replays the three that carry a replayable prefix (four failure
points), and one candidate is stopped at the gate (Section~\ref{sec:spend}).}
\label{tab:e2rate}
\begin{tabular}{@{}rrr@{}}
\toprule
Rewrites allowed & Repaired & Rate \\
\midrule
0 & 0/8 & 0\% \\
1 & 7/8 & 87.5\% \\
2 & 8/8 & 100\% \\
3 & 8/8 & 100\% \\
\midrule
replay-tested by E2b & 3/8 & --- \\
\bottomrule
\end{tabular}
\end{table}

All eight repairs pass the checker. Mean rewrites on successful tasks are 1.12, the first-attempt rate is 7 of 8 with a Wilson 95\% interval of [0.53, 0.98], and the rate within two rewrites is 8 of 8 with [0.68, 1.0]. The checker intercepts 53\% of candidates before
acceptance. All eight repaired scripts preserve the capability-call multiset, and three
contain no capability call, so the multiset guard applies to the other five, which preserve
it.

The passing edits are local, a bound or a slice, except for \texttt{a425}, which moves a
committed call into a helper and is stopped at replay (RQ1). The one two-rewrite case first
cleared R03 and R12 but introduced an R05 fanout-without-guard violation, and the second
candidate cleared both, which is why this sample averages 1.12.

\paragraph{A 32-script benchmark (E2-32)}
\label{sec:e2scale}

E2 has eight failures, so we drew 32 unique model-authored scripts from the E1 outputs that
fail the current checker. Twenty-two come from Gemini-3.5-Flash and ten from
DeepSeek-V4.1-Flash. All are loop-bound failures, six also carry R05, and all stay within the
four production-bound capabilities. With a two-rewrite budget the loop repairs 32 of 32 at the
first rewrite, with a mean of 1.00. All 32 preserve the call multiset, and the checker
intercepts half of the candidates.

The 32 repairs are generated before the prefix is built, so the store is empty at repair time
and the model sees no frozen-site list. We then replay each repaired script against the
maximal prefix built from its original script. Twenty-six carry a prefix that the production annotator compiles and that has at least two calls, and six have fewer than two calls. Of the 26, the
gate accepts 22 and stops four as \texttt{removed}. The 22 replay 258 committed identities,
execute 22 new calls, re-execute no committed identity, and report three unused committed
occurrences as orphans. The scale set reproduces the E2b result, and the gate separates static repair success from replay success.

\begin{table}[t]
\centering
\footnotesize
\setlength{\tabcolsep}{4pt}
\caption{Repair at scale. The 32-failure benchmark is drawn from E1 model outputs that fail
the current checker (loop-bound failures, and six also carry R05). Historical E2 replay is four
points on three scripts. The benchmark replay is twenty-six scripts.}
\label{tab:e2scale}
\begin{tabular}{@{}lrr@{}}
\toprule
 & Historical (E2) & Benchmark (E2-32) \\
\midrule
Failures & 8 & 32 \\
Repaired at one rewrite & 7/8 & 32/32 \\
Repaired within budget & 8/8 & 32/32 \\
Mean rewrites & 1.12 & 1.00 \\
Call multiset preserved & 8/8 & 32/32 \\
Replay-tested & 4 points & 26 \\
Gate pass & 3/4 & 22/26 \\
Replayed identities & 8 & 258 \\
New executions & 4 & 22 \\
Committed identity re-executed & 0 & 0 \\
\bottomrule
\end{tabular}
\end{table}

\paragraph{Four rules}

The 32-failure benchmark isolates R03, so we built a 35-script corpus from the P2
destroying-mutant corpus to test rule generality. The rules are R03 with 9 scripts, R04 with
9, R05 with 6, and R12 with 11, each with at least two capability calls. All 35 repair at one
rewrite and 34 preserve the call multiset. Thirteen carry replayable prefixes, and 22 are
reported separately. The gate passes eight and stops five, and every stopped candidate is an
R05 repair that removes the unguarded parallel call, so the static repair succeeded while the
committed callsite disappeared. The eight passing scripts replay 148 committed identities with
no duplicate. The replay-tested R03, R04, and R12 repairs all pass, with 51, 52, and 45
identities.

\paragraph{Prompt ablation}
\label{sec:probes}

E2 reports a 53\% static interception rate, and to see which prompt changes that rate, we ran the
four E2b failure points under three prompts. Arm A gives the runtime error only, arm B adds
the frozen sites, and arm C adds to B an explicit prohibition on moving a committed call into
a helper. The rewriter is \texttt{gemini-3.5-flash} at \texttt{temperature=0.3}, and we run
four samples per arm.

The pooled result favors listing the prefix. Arm A takes 1.00 mean attempts and produces 3 accepted candidates out of 12, arm B takes 2.19 and produces 8 of 12, and arm C takes 2.00
and produces 9 of 12.

Arm A measures the unlisted-prefix condition, in which 1.00 attempts suffice across all four samples, because the model produces a checker-valid candidate in one rewrite, and three of twelve final candidates pass the gate. In that condition the default behavior of the model is to
restructure a committed call. Listing the frozen sites changes the result. Admissibility rises from 3 of 12 to 8 of 12 at a mean of 2.19 attempts, and the gate stops fourteen candidates along
the way that the model then repairs, while under arm A those candidates would have been the
final answer. Arm C reaches 9 of 12 at 2.00 attempts, within noise of arm B at this sample
size, so we keep B as the shipped prompt.

Pooling the three arms and three samples, 36 final candidates reach the gate, 20 pass and 16
stop, and every stop is a removed callsite. Across the 40 final candidates drawn from real
failures, which adds four arm-A candidates from the fourth sample, 23 are accepted and 17 stop, all for a moved or deleted callsite.

A gate alone turns a restructure into a retry, and a prompt alone has no enforcement. The
useful prompt change is naming what is frozen, and the pair works because the gate supplies a
counterexample and the prompt supplies a model that can read it.

\paragraph{A second model}

With \texttt{deepseek-v4.1-flash} the loop repairs 5 of 8 at one rewrite and 8 of 8 within three, with a mean of 1.38. For \texttt{gemini-3.5-flash} the rates are 7 of 8 and 8 of 8 within two, with a mean of 1.12. Bounded convergence holds for both models, and the first-attempt
rate differs.

\subsection{Portability (RQ4)}
\label{sec:cross-domain}

The media studies establish the mechanism on high-cost calls, so we built twelve
customer-support tasks with a different shape. A task classifies a ticket, extracts fields,
drafts a policy-compliant reply, and decides whether a human must review it. The tasks include
single tickets, three-ticket fanouts, candidate selection, policy-section summarization,
missing-information follow-up, and a policy-conflict escalation. The capability is
\texttt{text.generate}, and the calls are billed and read-only, so the mechanism is exercised
without a media artifact.

Two models, \texttt{gemini-3.5-flash} and \texttt{deepseek-v4.1-flash}, wrote scripts at
temperature zero. Table~\ref{tab:cross-domain} reports the study, in which 24 scripts, twelve per model, pass R00 to R12 and use the granted capability. Nine of the accepted scripts were
executed against the live text provider and scored by the deterministic rubric of the task,
and all nine pass. An author read of three outputs agrees with the rubric. The pass rate is
over the nine executed scripts, and the table reports the per-cell denominators.

\begin{table}[t]
\centering
\footnotesize
\setlength{\tabcolsep}{3pt}
\caption{Customer-support study. \emph{Chk.} counts checker-valid scripts, \emph{Exec.}
the scripts executed against the live text provider, \emph{Rub.} the rubric passes,
\emph{Rep.} the replay points, \emph{Ids} the replayed identities, and \emph{Dup.} the
committed identities executed again. The executed and replay subsets are fixed and reported
with per-cell denominators.}
\label{tab:cross-domain}
\begin{tabular}{@{}lrrrrrrr@{}}
\toprule
Model & Tasks & Chk. & Exec. & Rub. & Rep. & Ids & Dup. \\
\midrule
Gemini-3.5-Flash & 12 & 12 & 6 & 6 & 2 & 4 & 0 \\
DeepSeek-V4.1-Flash & 12 & 12 & 3 & 3 & 2 & 4 & 0 \\
\midrule
Total & 24 & 24 & 9 & 9 & 4 & 8 & 0 \\
\bottomrule
\end{tabular}
\end{table}

Four accepted scripts were then replayed after a tail failure injected at the third text call.
Eight committed identities were replayed and no committed identity was executed again. A support fanout probe tests the boundary, where a repair moves the committed \texttt{text.generate}
call into a helper, and the production gate stops it as \texttt{removed} before dispatch. The
study applies identity, replay, and the gate to a second domain, and the repair loop itself is
measured on the media corpus.

\subsection{Authoring and checking (RQ5)}
\label{sec:authoring}

RQ5 covers authoring and checking. We measure authoring on 60 tasks, execution on the accepted scripts, held-out transfer, task success with human scoring, and checker validity on two mutation corpora.

\paragraph{Authoring (E1)}
\label{sec:e1}

E1 contains 60 tasks, of which 22 are anchors from real batches and 38 are synthetic, and they cover nine capabilities, 49 tasks within the four production-bound capabilities, 34
\texttt{sandbox} tasks, and 26 \texttt{dry\_run} tasks. Each model receives the same skill
document, capability catalog, and fixed envelope, and reference scripts are not included.
Sampling uses \texttt{temperature=0}, one sample per task, and \texttt{max\_tokens=16384}, and
a sample is accepted when R00 to R12 pass.\footnote{Four DeepSeek samples were regenerated
with a 32,768-token budget.} E1 runs without a container or a provider, and the models are
\texttt{deepseek/deepseek-v4.1-flash} and \texttt{gemini/gemini-3.5-flash}.

\begin{table*}[t]
\centering
\small
\caption{E1 checker acceptance on 60 tasks per model.}
\label{tab:e1}
\begin{tabular}{@{}lrrrr@{}}
\toprule
Model & $n$ & Accepted & Rate & 95\% Wilson \\
\midrule
DeepSeek-V4.1-Flash & 60 & 59 & 98.3\% & [91.1, 99.7] \\
Gemini-3.5-Flash & 60 & 51 & 85.0\% & [73.9, 91.9] \\
Pooled & 120 & 110 & 91.7\% & [85.3, 95.4] \\
\bottomrule
\end{tabular}
\end{table*}

Table~\ref{tab:e1} reports 110 of 120 checker accepts. Ten samples fall outside the checker's model, and all are R03. Manual review finds eight with a static cap outside the checker
model and two without a cap, so 110 of 120 is a lower bound on acceptance. Capability matching
is higher, at 117 of 120, so naming the right capabilities is easier than satisfying every
syntactic obligation. The difference between 59 of 60 and 51 of 60 is significant on the 60 paired tasks (McNemar \(p=0.02\)), and it describes this sample. The four regenerated samples turn the original 106
accepted outputs into 110, and E1b-60 uses the original 106.

A pre-registered 20-task subset is sampled three times at temperature 1.0. Gemini passes 17 of 20, 16 of 20, and 20 of 20, and DeepSeek passes 17 of 20, 15 of 20, and 19 of 20, with a sample standard deviation of about 10 points each. Single-sample acceptance rates carry sampling
variance, and we report the two temperature settings separately.

\paragraph{Execution (E1b)}

Static acceptance is the first gate and execution is the second. We pre-registered a 24-task
stratified subset with 12 production-bound, 6 sandbox, and 6 dry-run tasks, and executed the
scripts accepted in the original E1 run against a deterministic local provider without
re-sampling a model. The local runtime uses the container \texttt{Context} and a local RPC
that enforces grants, approvals, provider-call and concurrency limits, and input schema, and
it runs without Docker or a provider. Of 48 nominal task-model units, 43 scripts were
executable.

All 43 terminated without an uncaught exception or a timeout, made no denied or over-budget
call, and returned a serializable object. They make 244 local calls, with a median of 4\,ms
and a p95 of 15.7\,ms, and 60\,KB of output. One adequacy boundary appears in Gemini
\texttt{e1-002}, which hard-codes \texttt{MAX\_SHOTS = 5} while the fixture has six shots, so
it makes 11 calls where 13 are required, with five image and five review calls instead of six
each. The gap is in adequacy, and the checker and runtime behave as specified.

A pre-registered 60-task extension reuses the 106 stored outputs accepted in the original run.
The frozen 24 retain their assertions, and the additional 36 use generic structural evidence,
which is a non-empty serializable result, no failure status, and at least one artifact per
expected capability. Semantic quality is measured separately in R3, and execution passes 97 of 106 (91.5\%, 95\% Wilson [84.6, 95.5]), the frozen subset reproduces 42 of 43, and the
extension contributes 55 of 63. Nine execution differences remain, which are six missing
capability calls, two output-shape failures, and one over-budget call. The 106 scripts make 554
local calls, with a median of 3\,ms.

\paragraph{Held-out transfer (P1-C)}
\label{sec:p1c}

The task set of E1 was used during checker development, so we pre-registered 20 new tasks
after freezing the definition and the execution manifest. The set covers approval and budget
boundaries (6), parallel and partial-failure handling (4), long loops and large inputs (4),
data-bearing payloads (3), and prompt-injection and privilege prompts (3). Two models run once
per task at temperature 0, which is 40 calls, and static-valid outputs run against the same
local provider.

Gemini accepts 17 of 20 and DeepSeek 18 of 20, for a descriptive pool of 35 of 40 (87.5\%).
Five fall outside the checker model, of which four are bounded idioms and one has two input-derived loops with no static cap. We keep 35 of 40 as the pre-registered lower bound. All 35
static-valid scripts pass the local provider, with no false pass.

\paragraph{Task success (R3)}

We ran twenty outputs end to end on real providers, four image tasks on Seedream and sixteen
text or storyboard tasks on Gemini. A task-specific rubric checks the requested artifact,
which covers shot counts and durations, candidate counts, review counts, coverage, and failure
groups. Eighteen of twenty pass, ten of ten authored and eight of ten repaired. One case is a
coverage boundary, where the repaired feedback workflow counts 9 of 20 items, and in one
repair the failure branch was not reached because the provider succeeded.

Three annotators independently scored the same twenty outputs, and a task passes when its mean
over the three dimensions is at least 3. All three pass 16 of 20 and agree on every task-level
verdict, with a Fleiss kappa of 1.0. Pairwise agreement ranges from 50 to 55 of 60 dimension
scores, with 156 of 180 exact (86.7\%), and no pair differs by more than one point on any
score. Their disagreements with the rubric are identical on six cases. Four are cases where the rubric passes an output that all three annotators judge below the bar. Two are cases where the rubric flags an output that all three accept, namely the coverage miss and the failure branch that the provider did not reach. The rubric is a floor for task completion, and the human read
is the stricter measure.

\paragraph{Checker validity (P2)}
\label{sec:p2}

Static acceptance is a syntactic verdict, so we measured both error directions on a
pre-registered mutation corpus. The 40 bases are a deterministic hash-selected subset of the
106 unique static-valid scripts of the original E1 run. Destroying mutants cover nine contract
rules, and preserving mutants cover five semantics-preserving edits.

The mutation study found and closed an R05 coverage gap. Indirect fanout was outside R05 when
the fanout function was reached through a list variable, while the direct form
\texttt{ctx.parallel([f(i) ...])} was rejected. After the fix, the 419-mutant corpus gives
\(\alpha=0/269\) with a 95\% Wilson interval of [0.0, 1.4] and \(\beta=0/142\) with [0.0, 2.6].
Every destroying operator hits its target rule, and the per-operator denominators are in the
artifact. The \texttt{missing\_review} layer, with 8 of 8 accepted, measures task-level
adequacy, which \(\alpha\) leaves outside its scope. Mutation testing provides a controlled
proxy \cite{JiaHarman11,Just14}, and our operators are rule-derived, so \(\alpha\) is a
conservative estimate for real contract failures. A zero rate describes these mutants, and the
runtime rule carries the safety claim. The Wilson intervals are descriptive because mutants
from one base are dependent.

On 35 held-out P1-C bases with 323 mutants, of which 205 are destroying, 111 preserving, and 7
semantic, the corpus gives \(\alpha=0/205\) with [0.0, 1.8] and \(\beta=0/111\) with [0.0, 3.3]. Seven \texttt{missing\_review} mutants remain outside \(\alpha\).

\subsection{Reliability and overhead (RQ6)}
\label{sec:reliability}

RQ6 covers the reliability and overhead of the machinery. We measure the crash window,
store failures, orphan reporting, and the latency of the gate.

\paragraph{The crash window (E4)}
\label{sec:e4}

E2b injects a handler exception and E3 raises inside the script. The fail-closed window needs
a durable attempt, a process that dies before commit, and a provider that may or may not have
completed. E4 creates that window with a real child process and the production SQLite store.
The child writes the attempt and then kills itself with SIGKILL. In the timeout case the parent kills a child whose handler sleeps. In the parallel case three threads write three attempts and synchronize on a barrier, and the process is killed. The parent then replays through \verb|_idempotent|.

Across the five cases, every unresolved attempt was detected before a handler ran, every
committed identity was replayed from the store, and every replay returned a stored result. After
\texttt{reconcile\_attempt(..., result=None)} the abandoned identity executes once and a
second replay is a cache hit. After \texttt{reconcile\_attempt(..., result=...)} the replay
returns the stored result and executes zero times. The parallel case leaves three attempts and
blocks all three, which is the \(k=3\) instance of the analytical bound. Whether the provider
completed before the crash is the uncertainty that the fail-closed protocol and reconciliation
exist for.

\paragraph{Store failure and orphans}

Storage errors are fail-closed by default, so a failed prefix listing or idempotency read raises \texttt{StoreUnavailableError} and the failure is visible. With a store that raises on every
read, a one-call script raises with zero handler executions, which trades one visible failure
for an invisible duplicate charge.

Shrinking a committed loop from three iterations to one leaves every frozen callsite in the
source, so the gate accepts, and only occurrence zero is invoked. Two committed identities
become orphans and the run otherwise succeeds. \texttt{DYNAMIC\_ORPHAN\_POLICY} is
\texttt{warn}, which persists the report, or \texttt{stop}, which persists the report and
fails the run. We keep \texttt{warn} as the default because an orphan is a fidelity event, and
a deployment that bills per artifact chooses \texttt{stop}.

\paragraph{Overhead}
\label{sec:cost}

The checker, the prefix gate, and the certificate stay in the low-millisecond range on one
host. The checker has a p50 of 1.70\,ms and a p95 of 4.21\,ms over 116 scripts and 2320
samples. The gate ranges from 0.041\,ms at one frozen site to 4.869\,ms at 200 frozen sites,
and certificate persist and verification stay below 0.55\,ms up to 200 sites. These are
in-process host-CPU numbers with an in-memory store, which is the algorithmic cost of the
mechanism. The artifact carries the per-bucket and per-size records.

\section{Threats to Validity}
\label{sec:threats}

\paragraph{Construct validity}

Our guarantee is identity-level, and it covers the calls that a candidate keeps and the
identities that it invokes. A new callsite that repeats an old provider request is outside it,
which the ledger reports and the policy may deny, and so is a duplicate inside the provider,
which needs a provider idempotency key. The crash window is bounded rather than closed. An
unresolved attempt fails closed, and an explicit abandonment can repeat at most \(k\) effects
per crash. The checker is both an instrument and an object of study, and its acceptance rate
describes the syntactic model of the contract rather than the semantic quality of a script,
which R3 measures separately. Because the mutation corpora are rule-derived, the zero
false-accept rate is a conservative estimate for real contract failures.

\paragraph{Internal validity}

Model-dependent measurements carry the pre-registered number and a separate variance study.
E1 samples once per task at temperature 0, while a 20-task subset sampled three times at
temperature 1.0 has a standard deviation of about 10 points, so a single-sample rate carries
sampling variance. The prompt ablation compares three prompts at four samples per arm, which is
enough to see a large effect and not enough to rank arms B and C. Instructed edits drive the
payload-edit study, so it shows that the model can produce the case and not how often it would
do so unprompted. Stored model outputs drive E2 and E7, which removes run-to-run variance at
the cost of describing one sampling round.

\paragraph{External validity}

One platform, one run store, and one sandbox image support the evaluation, and the media domain
dominates it. The customer-support study broadens the task distribution with a read-only,
lower-cost capability, and it executes 9 of its 24 scripts. Two billed image providers, one
text provider, two model families, one full durable engine, and one step-fork probe cover the
provider and engine axes at the resolution of a case study. A deployment with hundreds of
capabilities, provider-side idempotency keys, or multi-host storage is outside the measured
set. The repair corpora are small by design, because each case needs a replayable prefix and
real provider failures are the scarce input.

\paragraph{Conclusion validity}

For the identities that were replayed, the zero duplicate count is exact, and the acceptance,
repair, and pass rates carry Wilson intervals with their denominators. Mutants from one base
share its structure, so the mutation corpora are not independent samples and those intervals
are descriptive, with per-operator denominators in the artifact. The 258 replayed identities
come from 22 scripts, and the eight real repairs are a strict subset of 32 checker failures.
Because the spend advantage is a property of the fanout shape and the sequential case ties, the
numbers should not be read as an average saving over repairs in general.

\section{Related Work}
\label{sec:related}

\subsection{Durable Execution and Replay}

Durable execution engines record a history and require the replayed code to reproduce its
command sequence. A tail-compatible edit replays, and an incompatible one becomes a
non-determinism failure or needs a version marker \cite{Temporal}. Both families make the same engineering choice. DAG orchestrators serialize a task graph and rerun the graph
\cite{Airflow24}, while durable runtimes journal each step and re-execute the code around the
journal \cite{DBOS25,Restate25}. Both replay a fixed program, and the source-level question
stays open.

\emph{Resume Means Resume} \cite{Resume26} formalizes prefix continuation, effect
exactly-once, and consume-once for a fixed program, and it excludes cross-version checkpoint
migration. \emph{LIMBO} \cite{Limbo26} proves that no verification-only policy is exactly-once
under late commits without a bound on in-flight time, and it measures that attaching
idempotency keys lowers the duplicate rate from 28\% to 4\%. \emph{State Without a Landlord}
\cite{State26} records \texttt{(callSite, callIndex, argsHash)} in replay capsules and assumes
the source unchanged. Compensation and checkpointed stream processing
\cite{Sagas87,ChandyLamport85,Flink15}, content-addressed build caches and dynamic software
updating \cite{Hicks05,Neamtiu06,BuildSystems18}, and provenance \cite{Davidson08} keep or
restore a history. None of them defines which source edits are allowed.

\subsection{Identity, Revision, and Repair}

The orchestration layer is increasingly a model artifact. ReAct \cite{ReAct23} interleaves
reasoning with tool calls, and AutoGen \cite{AutoGen23} and LangGraph \cite{LangGraph24}
compile multi-agent interaction into stateful graphs. These frameworks make the plan flexible
and leave open which part of a rewritten graph may still touch a paid effect. \emph{Pollard} \cite{Pollard26} assigns semantic step identities over
caller-selected canonical input, kind, attempt, and parent. Its strict replay fails closed, and its record and hybrid modes execute the new step. The identity is semantic and chosen by the caller, while ours is syntactic and computed by the host, which is what lets a pre-spend check run without the model's cooperation. \emph{ControlScope} \cite{ControlScope26}
compares keeping the code, editing the arguments of the next call, and replacing the unfinished
continuation, and it measures task success. Recent work repairs authorization \cite{Apas26},
pending decisions \cite{Atr26}, and rollback \cite{SafeResume26}, and it studies multi-agent
plans and handoffs \cite{Alas25,Bridge26} and gated or verified execution
\cite{SkillNB26,Tessara26}. Classical instance migration \cite{Rinderle03,Rinderle04,Song19} is
the closest historical analogue, and it can discard earlier loop iterations and recompute a
marking. A paid provider call cannot be discarded that way.

\paragraph{Repair loops and effect validation}
LLM repair is usually framed as patch generation. Reflexion \cite{Reflexion23} and
Self-Refine \cite{SelfRefine23} iterate on feedback, ChatRepair \cite{ChatRepair24} repairs
bugs through conversation, and the APR literature \cite{APRsurvey19} judges a patch by whether
it passes tests. Our loop also uses structured counterexamples, but the acceptance
predicate is different. A candidate is admitted only if it preserves the call expressions that
produced committed effects. A patch can pass every test and still move a committed call, and
the gate stops it before the provider is paid.

\paragraph{Three adjacent lines of work}
Workflow instance migration \cite{Rinderle03,Rinderle04,Song19} has structural, behavioral, and
dynamic correctness criteria, but it can roll back a marking and recompute state. A paid
provider call cannot be recomputed for free, so the contract adds a pre-spend source predicate
rather than a migration rule. A content-addressed cache or a provider idempotency key
\cite{StripeIdem} removes duplicate charging at the provider, but it does not decide which source edits are allowed, and it can return the artifact of another callsite, which RQ2
measures. The contract composes with a provider key and does not replace it. APR patch
validation asks whether a patch is correct, while the gate asks whether a revision may still
touch a committed effect. The two predicates are independent. Our candidate must pass the checker and the gate, and the gate can stop a patch that is semantically correct.

\subsection{Positioning}

\begin{table*}[t]
\centering
\footnotesize
\caption{Nearest neighbors by the question the revision contract has to answer. The two
middle columns are the mechanically checkable ones, because they ask whether the system
defines a source-level set that a rewrite may touch and whether it stops a rewrite before a paid effect.}
\label{tab:neighbors}
\begin{tabular}{@{}>{\raggedright\arraybackslash}p{2.2cm}>{\raggedright\arraybackslash}p{3.0cm}>{\raggedright\arraybackslash}p{2.5cm}>{\raggedright\arraybackslash}p{2.4cm}>{\raggedright\arraybackslash}p{5.2cm}@{}}
\toprule
Work & Unit & Writable set & Pre-spend reject & What the contract adds \\
\midrule
Resume \cite{Resume26} & fixed program + persistence API & no (fixed program) & yes (non-determinism) & source-level revision across versions \\
Pollard \cite{Pollard26} & semantic step identity & no (identity only) & yes (hash seal) & a writable set with a source predicate \\
ControlScope \cite{ControlScope26} & runtime revision scope & no (runtime scope) & -- & a committed-effect set \\
State \cite{State26} & replay capsule & no (source assumed unchanged) & no & revision over a replay capsule \\
APAS \cite{Apas26} & authorization-bearing action & no & yes (authorization) & source callsites as the unit \\
ATR \cite{Atr26} & pending decision & no & yes (pre-effect) & source rewrite as the unit \\
SafeResume \cite{SafeResume26} & checkpoint/rollback & no (backward) & no (rollback) & forward revision after a tail failure \\
SKILL.nb \cite{SkillNB26} & reusable workflow artifact & partial & yes (gate) & one partial execution's committed boundary \\
ALAS \cite{Alas25} & multi-agent plan & no (log level) & partial & callsite identity and a committed boundary \\
BRIDGE \cite{Bridge26} & multi-agent handoff & no & partial & source callsites \\
\midrule
Temporal \cite{Temporal} & command sequence & partial (tail-compatible) & no (replay-time) & a pre-spend source predicate and a payload digest \\
Ours & callsite $(\sigma,\pi,\omega)$ & \textbf{yes} (append, site-level freeze) & \textbf{yes} (static gate) & the full column, with writable set, gate, and digest \\
\bottomrule
\end{tabular}
\end{table*}

Table~\ref{tab:neighbors} states the difference by system. Identity, authorization, and lifecycle
are established machinery, and the source-level writable set, the callsite-level freeze, and
the pre-spend predicate are the additions of the contract.

The identity is established machinery, and what the contract adds is the allowed set and its measured limit. We define which source edits preserve the committed calls, establish the soundness of
the gate and describe its cost on relocation, and measure the limit against a durable engine.
The spend bound follows, and it is strict on partial fanout.

\section{Conclusion}
\label{sec:conclusion}

A workflow written by a model can be revised without paying twice for what it already did, and
without returning a result that the revision has replaced. The contract gives every capability call a structural identity and freezes the call expression that produced a committed payload. It stops a rewrite that moves or changes that expression before a container starts, and it stops a committed payload change at dispatch with a digest.

The gate stops a relocation that preserves the call, its payload, and its order, because under structural identity the moved call is new and would execute. Proposition~\ref{prop:dilemma} states why no local identity avoids the choice. The
digest closes the other case and fails closed when the past must change. Against a durable
engine's reset, the contract saves calls on a partial fanout and stops a payload change that
reset reuses, while it stops a relocation that reset completes and ties on sequential
repairs.

Two directions follow from the result. A pre-spend check that works on traces rather than
callsites could accept a relocation while preserving the committed call sequence, and the
evidence can be widened to more platforms, more domains, and more engines.

\bibliographystyle{IEEEtran}
\bibliography{references}

\end{document}